\documentclass[reprint, prd]{revtex4-2}
\usepackage{amsmath, amssymb}
\usepackage{graphicx,color,float,tabularx,siunitx}
\usepackage[caption=false]{subfig}
\usepackage{placeins}
\usepackage{rviewport}
\usepackage{orcidlink}
\definecolor{colorLink}{rgb}{0,0,180} 
\usepackage{hyperref}
\hypersetup{
   colorlinks = true,
   citecolor  = colorLink,
   urlcolor   = colorLink,
   linkcolor  = colorLink,
}

\DeclareSIUnit\year{yr}
\usepackage{subfig}
\usepackage{float} 

\usepackage[sort&compress]{natbib}

\usepackage[shortcuts]{extdash}

\begin{document}
\title{Probing Black Hole Phase Transitions through Quasi-Periodic Oscillations}

\author{Bidyut Hazarika \orcidlink{0009-0007-8817-1945}$^1$}

\email{rs\_bidyuthazarika@dibru.ac.in}

\author{Prabwal Phukon \orcidlink{0000-0002-4465-7974}$^1$,$^2$}
\email{prabwal@dibru.ac.in}

	\affiliation{$^1$Department of Physics, Dibrugarh University, Dibrugarh, Assam, 786004.\\$^2$Theoretical Physics Division, Centre for Atmospheric Studies, Dibrugarh University, Dibrugarh, Assam, 786004.\\}


\begin{abstract}
In this work, we probe the well known thermodynamic phase structure of black hole through the lens of its quasi-periodic oscillations (QPOs). Can QPOs be influenced by black hole phase transitions? Do they carry any signature of such transitions in their observational patterns? These were the central questions guiding our study. 
Using both RN AdS and Kerr black hole backgrounds across different QPO models,  we analyzed the behavior of upper and lower QPO frequencies as functions of the Hawking temperature.  Our results shows that  QPO frequencies trace out distinct thermodynamic phases and also reflect their stability properties.  As the black hole transitions between different thermodynamic phases, the trend of QPO frequencies with respect to temperature also shifts. Due to lack of of observational data, the present work is primarily more on the mathematical side,  as the underlying mechanism responsible for the Hawking temperature has not yet been fully understood or experimentally verified. Moreover, given the speculative nature of black hole phase transitions, it would be unfair to claim that our results establish a definitive connection between an observable quantity such as the QPO frequency and the thermodynamic phase behavior of black holes. Nevertheless, our analysis suggests a possibility that changes in black hole geometry could be one of the contributing factors influencing QPO behavior.

\end{abstract}


\maketitle                                                                      

\section{Introduction}
Black holes are among the most enigmatic and intriguing consequences predicted by General Relativity (GR), highlighting the theory's profound impact on our understanding of gravity. Since the formulation of GR by Einstein in 1915, it has served as a solid theoretical framework for describing the curvature of spacetime. A pivotal confirmation of GR came with the detection of gravitational waves from black hole mergers by the LIGO collaboration, a groundbreaking achievement in the field of gravitational physics \cite{ligo}. This was followed by the historic imaging of supermassive black holes by the Event Horizon Telescope (EHT), first capturing the black hole in the M87 galaxy, and later in Sagittarius A* (SgrA*) at the center of our own Milky Way \cite{m87a, m87b, m87c, m87d, m87e, m87f}. These revolutionary images unveiled a dark central region surrounded by a glowing photon ring, the characteristics of which provide valuable insights into the nature of black holes and the gravitational framework governing them \cite{Shadow1,Shadow2,Shadow3,Shadow4}.\\

In the early 1970s, pivotal research in black hole thermodynamics laid the foundation for understanding the connection between black holes and the laws of thermodynamics \cite{Bekenstein:1973ur,Hawking:1974rv,Hawking:1975vcx,Bardeen:1973gs}. This foundational work has since been expanded upon by numerous studies, revealing a wealth of fascinating insights into this field \cite{Wald:1979zz,bekenstein1980black,Wald:1999vt,Carlip:2014pma,Wall:2018ydq,Candelas:1977zz,Mahapatra:2011si}. 
One intriguing aspect of black hole thermodynamics is the exploration of phase transitions \cite{Davies:1989ey,Hawking:1982dh,curir_rotating_1981,Curir1981,Pavon:1988in,Pavon:1991kh,OKaburaki,Cai:1996df,Cai:1998ep,Wei:2009zzf,Bhattacharya:2019awq,Kastor:2009wy,Dolan:2010ha,Dolan:2011xt,Dolan:2011jm,Dolan:2012jh,Kubiznak:2012wp,Kubiznak:2016qmn}. Black holes can exhibit various types of phase transitions, including the Davies type phase transition \cite{Davies:1989ey}, the Hawking–Page phase transition \cite{Hawking:1982dh}, the extremal phase transition (which describes the transition of black holes from non-extremal to extremal states) \cite{curir_rotating_1981,Curir1981,Pavon:1988in,Pavon:1991kh,OKaburaki,Cai:1996df,Cai:1998ep,Wei:2009zzf,Bhattacharya:2019awq}, and the Van der Waals-type phase transition (which is similar to the phase transition seen in Van der Waals fluids) \cite{Kastor:2009wy,Dolan:2010ha,Dolan:2011xt,Dolan:2011jm,Dolan:2012jh,Kubiznak:2012wp,Kubiznak:2016qmn}.\\

Quasiperiodic oscillations (QPOs), detected in the X-ray flux from black holes and neutron stars, have emerged as powerful tools for probing the physics of strong gravity regimes. Characterized by brightness fluctuations at nearly periodic intervals, these oscillations are widely believed to originate from fundamental processes within the accretion disk, including relativistic effects and complex disk dynamics. In particular, the observation of twin-peak QPOs in several compact object systems has spurred extensive theoretical and observational studies, often attributing these features to resonances or intrinsic oscillatory behavior within the disk structure. As observational techniques continue to improve, the demand for more sophisticated theoretical frameworks becomes increasingly evident. Since their first identification through spectral and temporal studies in X-ray binaries \cite{17}, QPOs have been examined from a range of perspectives.
Among various theoretical models, those focusing on the motion of particles in curved spacetime have gained significant attention. In such scenarios, the modulated trajectories of charged test particles are considered central to shaping the accretion flow and driving the observed oscillatory signals \cite{18,19,20,21,22,23,24,25,26,27,28,29,30,31,32}. Recent computational efforts have employed general relativistic hydrodynamic simulations to explore QPO formation mechanisms around black holes, particularly in spacetimes such as Kerr and hairy black holes \cite{33}. These studies suggest that perturbations in the accreting plasma can generate spiral shock patterns, which are intimately connected to the emergence of QPOs \cite{34,35,36}. Likewise, models based on Bondi–Hoyle–Lyttleton accretion dynamics show that shock cones, formed due to the gravitational influence of compact objects, can produce distinct QPO signatures \cite{37,38,39,40,41}.
Such approaches have not only succeeded in reproducing the QPO characteristics of well-known sources like GRS 1915+105 \cite{42}, but also offer predictive insight into oscillatory behavior near supermassive black holes, including the one in M87 \cite{43}. The dynamics of test particle motion and the resulting QPOs around black holes remain a rich area of investigation key contributions can be found in Refs.\cite{c1,c2,c3,c4,c5,c6,c7}.\\

The objective of this work is to explore possible imprints of black hole thermodynamic properties using quasi-periodic oscillations (QPOs).  As a standard practice we start with the Reissner-Nordström Anti-de Sitter (RN-AdS) black hole.  Although the RN-AdS black hole is not directly representative of astrophysical black holes due to its static nature and asymptotically non-flat boundary but it serves as an excellent theoretical model for exploring black hole phase transitions.  In particular, the RN-AdS geometry possesses a well-understood phase structure, including Van der Waals-like transitions and clear thermodynamic branches, which makes it an ideal testbed for probing whether quasi-periodic oscillation (QPO) frequencies can encode information about underlying black hole phase transitions. The aim of this study is thus conceptual : to investigate if QPO signatures can reflect thermodynamic phenomena in black hole spacetimes.The background geometry will be upgraded to the Kerr spacetime later in this work. Since Kerr black holes are rotating and asymptotically flat, they more accurately describe observed black hole systems, particularly those involving accretion and QPO generation.  It is important to mention that,  in this work, we focus solely on the relationship between Hawking temperature and QPO frequencies, treating Hawking temperature as a theoretical control parameter reflecting the black hole's thermodynamic state. We do not incorporate the effects of the temperature of accretion disk in our analysis. 

\section{RN $AdS$ black hole}
\subsection{Thermodynamics of RN AdS black hole}
The static, spherically symmetric metric of RN AdS black hole is of  the form
\begin{equation}
ds^2 = -f(r) dt^2 + \frac{dr^2}{f(r)} + r^2 \left( d\theta^2 + \sin^2\theta\, d\phi^2 \right),
\end{equation}
with the lapse function \( f(r) \) given by
\begin{equation}
f(r) = 1 - \frac{2M}{r} + \frac{Q^2}{r^2} + \frac{r^2}{\kappa^2},
\end{equation}
We set the AdS length $\kappa=1$ for the rest of our analysis. Here \( M \) and \( Q \) represent the ADM mass and electric charge of the black hole, respectively.

The mass parameter \( M \) can be expressed in terms of the event horizon radius \( r_+ \) as
\begin{equation}
M = \frac{Q^2+r_+^4+r_+^2}{2 r_+}
\end{equation}
The temperature of the black hole can be calculated by using the formula,
\begin{equation}
T = \frac{1}{4 \pi} f'(r)_{r=r_+}=\frac{-Q^2+3 r_+^4+r_+^2}{4 \pi  r_+^3}
\end{equation}
The critical values associated with the system can be found by solving the following equations
\begin{equation}
\frac{d T}{d r_+}=0 \quad ,\quad \frac{d^2 T}{d r_+^2}=0
\end{equation}
The solution of the thermodynamic equations reveals the existence of a critical charge given by
\begin{equation}
Q_C = \frac{1}{\sqrt{6}},
\end{equation}
below which (\( Q < Q_C \)) the RN-AdS black hole exhibits a Van der Waals (VdW)-like phase transition. In this regime, the system displays three distinct black hole phases: a small black hole (SBH) branch, an intermediate black hole (IBH) branch, and a large black hole (LBH) branch. For \( Q > Q_C \), the phase transition behavior disappears. The local thermodynamic stability of these black hole phases can be analyzed through the sign of the specific heat, defined by
\begin{equation}
C = \frac{dM}{dT},
\end{equation}
where \( M \) is the mass and \( T \) is the Hawking temperature of the black hole. A positive specific heat indicates local stability, while a negative value implies instability. To examine global thermodynamic stability, one typically analyzes the Helmholtz free energy. However, as these features of RN-AdS black holes are well-established in the literature, we summarize the key outcomes: the SBH and LBH branches are locally thermodynamically stable due to their positive specific heat, whereas the IBH branch is unstable, as it is characterized by a negative specific heat.\\

In this work, we aim to investigate the potential connection between the thermodynamic phase transitions of  black holes and the behavior of their associated quasi-periodic oscillations (QPOs). Our primary objective is to examine whether QPOs can serve as reliable indicators of black hole phase structure. To this end, we begin our study by evaluating the QPO frequencies within the Relativistic Precession (RP) models.  Our analysis involves considering both massless and massive particles in orbit around the black hole. We extend our analysis to Warped Disk (WD), and Epicyclic Resonance (ER) frameworks also. The results are being presented concisely in the conclusion section. 

\subsection{QPO and phase transition}
\begin{figure*}[!t]
\centering
\includegraphics[width=0.45\linewidth]{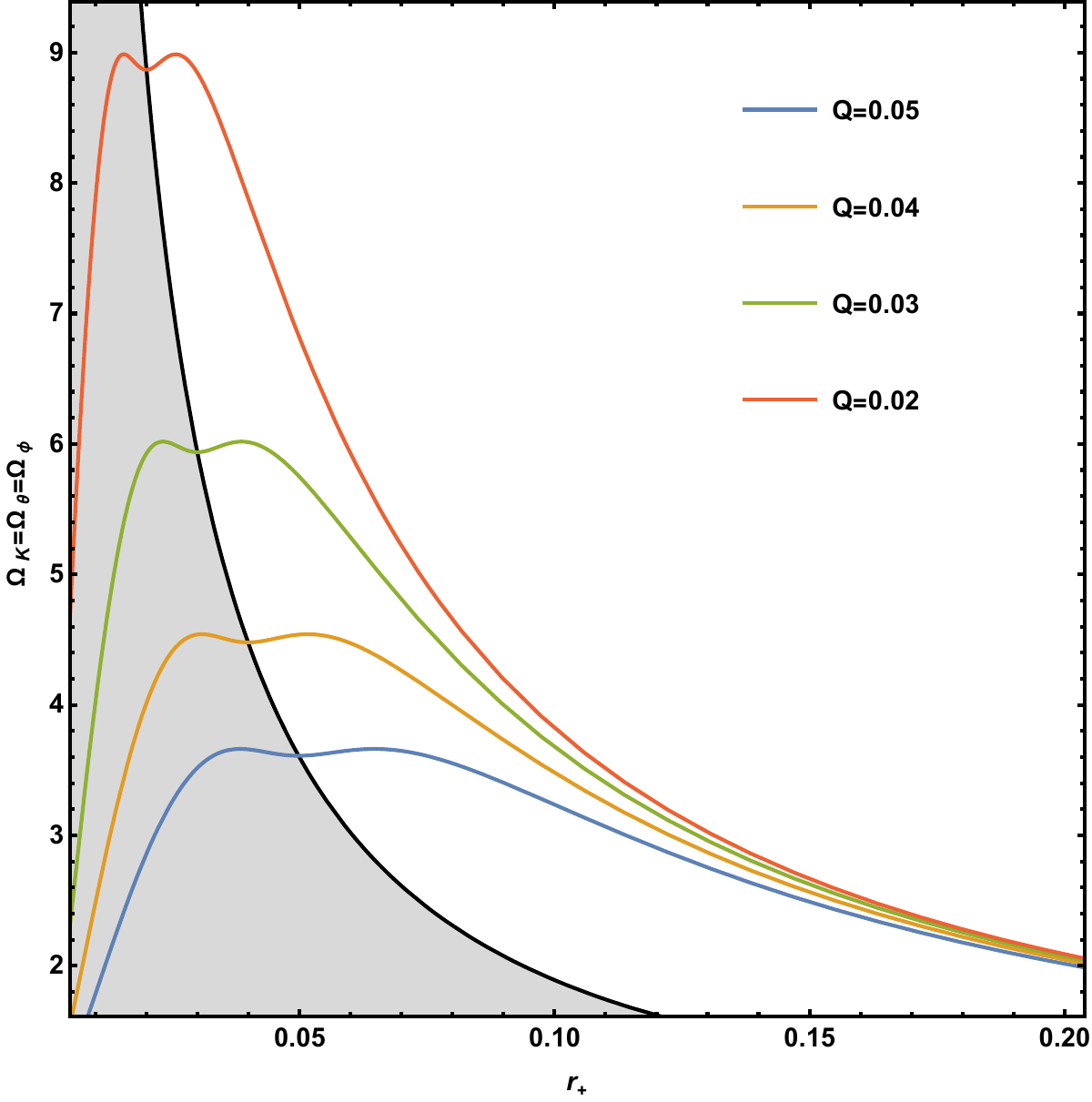}  
\hspace{0.5cm}
\includegraphics[width=0.45\linewidth]{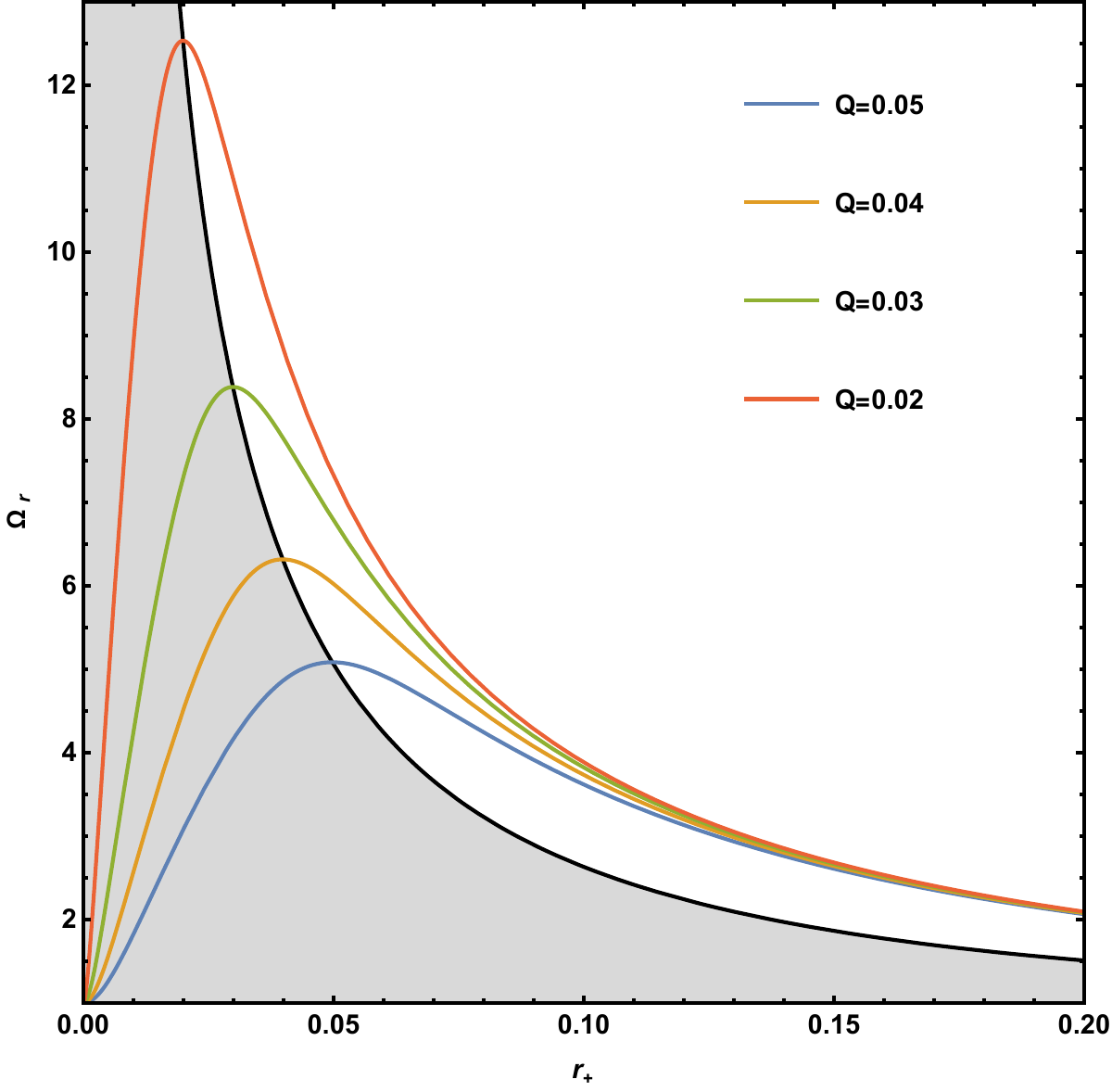}
\vspace{-0.3cm}
\caption{Radial and vertical oscillatory response of the particle in a unstable circular null geodesics with different values of $Q$.  Gray area represents the non-physical region where temperature is negative.  Both axes are expressed in natural units.
}
\label{fig1}
\end{figure*}
\label{LyapunovExponent}
We begin by considering the Lagrangian for a test particle constrained to the equatorial plane, i.e., $\theta = \pi/2$. The Lagrangian is given by:
\begin{equation}
\label{Eq:Lagrangian_Rephrased}
\mathcal{L} = \frac{1}{2} \left[ -f(r)\left(\frac{dt}{d\tau}\right)^2 + \frac{1}{f(r)}\left(\frac{dr}{d\tau}\right)^2 + r^2 \left(\frac{d\phi}{d\tau}\right)^2 \right],
\end{equation}
where $\tau$ denotes the proper time and $f(r)$ is the lapse function.

The canonical momenta associated with the generalized coordinates are computed as:
\begin{align}
\label{Eq:Canonical_Momenta_Rephrased}
p_t &= \frac{\partial \mathcal{L}}{\partial \dot{t}} = -f(r)\dot{t} = -E, \\
p_r &= \frac{\partial \mathcal{L}}{\partial \dot{r}} = \frac{1}{f(r)}\dot{r}, \nonumber \\
p_\phi &= \frac{\partial \mathcal{L}}{\partial \dot{\phi}} = r^2\dot{\phi} = L, \nonumber
\end{align}
where $E$ and $L$ represent the conserved energy and angular momentum of the particle, respectively. Here, a dot represents differentiation with respect to $\tau$.

Solving these equations yields the velocities:
\begin{equation}
\label{Eq:Velocities_Rephrased}
\dot{t} = \frac{E}{f(r)}, \quad \dot{\phi} = \frac{L}{r^2}.
\end{equation}

We then proceed to compute the Hamiltonian using the Legendre transformation:
\begin{equation}
\label{Eq:Hamiltonian_Rephrased}
\begin{aligned}
2\mathcal{H} &= 2(p_t \dot{t} + p_r \dot{r} + p_\phi \dot{\phi} - \mathcal{L}) \\
&= -f(r)\dot{t}^2 + \frac{\dot{r}^2}{f(r)} + r^2\dot{\phi}^2 \\
&= -\frac{E^2}{f(r)} + \frac{\dot{r}^2}{f(r)} + \frac{L^2}{r^2} = -\delta_1,
\end{aligned}
\end{equation}
where we have used Eq.~\eqref{Eq:Velocities_Rephrased}. Here, $\delta_1 = 1$ corresponds to timelike geodesics, and $\delta_1 = 0$ corresponds to null geodesics.

Introducing the effective potential $V_r$ for the radial motion through the relation $V_r = E^2-\dot{r}^2$, we obtain:
\begin{equation}
\label{Eq:m}
V_r = f(r)\left[\delta_1 + \frac{L^2}{r^2} \right].
\end{equation}

\subsection{Massless particle (null geodesic) : RN black hole case}
To study the QPO for massless particle we set $\delta_1=0$. The condition for unstable geodesic is $V'_r(r_0)=0$ and $V''_r(r_0)<0$. From \eqref{Eq:m} we can have
\begin{equation}
V'_r(r_0)=\frac{L f'(r_0)}{r_0^2}-\frac{2 L f(r_0)}{r_0^3}=0
\label{m2}
\end{equation}
From Eq. \eqref{m2} it is evident that $r_0$ is independent of L
Using the lapse function of RN AdS black hole,  Eq.\eqref{m2} is solved for $r_0$ .  Now substituting the expression of mass in terms of event horizon radius $r_+$,  we obtained the following relation between radius of the unstable circular orbit  $r_0$ and  event horizon radius $r_+$ as 
\begin{equation}
r_0=\frac{1}{2} \left(\sqrt{\frac{9 \left(Q^2+r_+^4+r_+^2\right){}^2}{4 r_+^2}-8 Q^2}+\frac{3 \left(Q^2+r_+^4+r_+^2\right)}{2 r_+}\right)
\end{equation}

To analyze the fundamental frequencies associated with oscillatory motion of test particles around a RN AdS black hole, one considers small deviations from a unstable circular orbit. Specifically, the coordinates are perturbed as \( r \to r_0 + \delta r \) and \( \theta \to \theta_0 + \delta \theta \), where \( r_0 \) and \( \theta_0 \) define the equilibrium position. The effective potential \( V_{\text{eff}}(r, \theta) \) can then be expanded in a Taylor series around this equilibrium configuration:

\begin{align}
V_{\text{eff}}(r, \theta) &= V_{\text{eff}}(r_0, \theta_0) + \delta r \left. \frac{\partial V_{\text{eff}}}{\partial r} \right|_{r_0, \theta_0} + \delta \theta \left. \frac{\partial V_{\text{eff}}}{\partial \theta} \right|_{r_0, \theta_0} \nonumber \\
&+ \frac{1}{2} \delta r^2 \left. \frac{\partial^2 V_{\text{eff}}}{\partial r^2} \right|_{r_0, \theta_0} + \frac{1}{2} \delta \theta^2 \left. \frac{\partial^2 V_{\text{eff}}}{\partial \theta^2} \right|_{r_0, \theta_0} \nonumber \\
&+ \delta r \delta \theta \left. \frac{\partial^2 V_{\text{eff}}}{\partial r \partial \theta} \right|_{r_0, \theta_0} + \mathcal{O}(\delta r^3, \delta \theta^3).
\label{exp}
\end{align}
At the circular orbit, the first-order derivatives of the potential vanish due to equilibrium conditions, and the higher-order terms are neglected under the small perturbation assumption. As a result, the equations of motion for the radial and vertical perturbations reduce to those of a harmonic oscillator. These perturbations manifest as oscillations with characteristic frequencies as measured by a distant observer \cite{Bambi2017book}:

\begin{align}
    \frac{d^2 \delta r}{dt^2} + \Omega_r^2 \delta r = 0, 
    \quad 
    \frac{d^2 \delta \theta}{dt^2} + \Omega_\theta^2 \delta \theta = 0,
\end{align}

where the squared radial and vertical frequencies are given by \cite{50}

\begin{align}
    \Omega_r^2 = -\frac{1}{2 g_{rr} \dot{t}^2} 
    \left. \frac{\partial^2 V_r}{\partial r^2} \right|_{r_0, \theta = \pi/2},
\end{align}
\begin{align}
    \Omega_\theta^2 = -\frac{1}{2 g_{\theta\theta} \dot{t}^2} 
    \left. \frac{\partial^2 V_r}{\partial \theta^2} \right|_{r_0, \theta = \pi/2}.
\end{align}
\begin{align}
    \Omega_\phi^2 = \frac{-\partial_r g_{tt}}{\partial_r g_{\phi\phi}} = \sqrt{\frac{f'(r)}{2r}}.
\end{align}
Here $ \Omega_\phi=  \Omega_\theta=  \Omega_K$.  $ \Omega_K$ is the Keplerian frequency which is the angular velocity of a test particle revolving around a black hole, as observed by a distant observer.
\begin{figure*}[t!]
\centering
\includegraphics[width=0.33\linewidth]{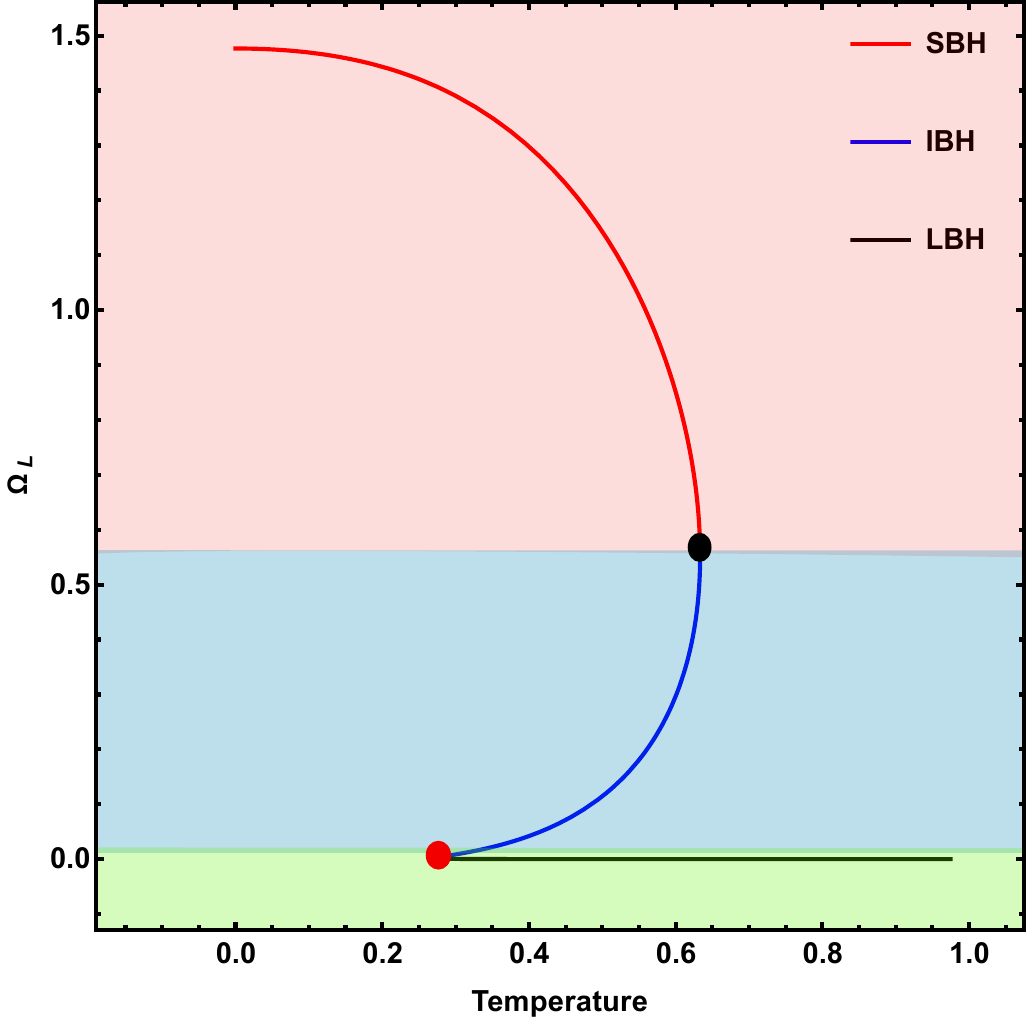}  
\includegraphics[width=0.32\linewidth]{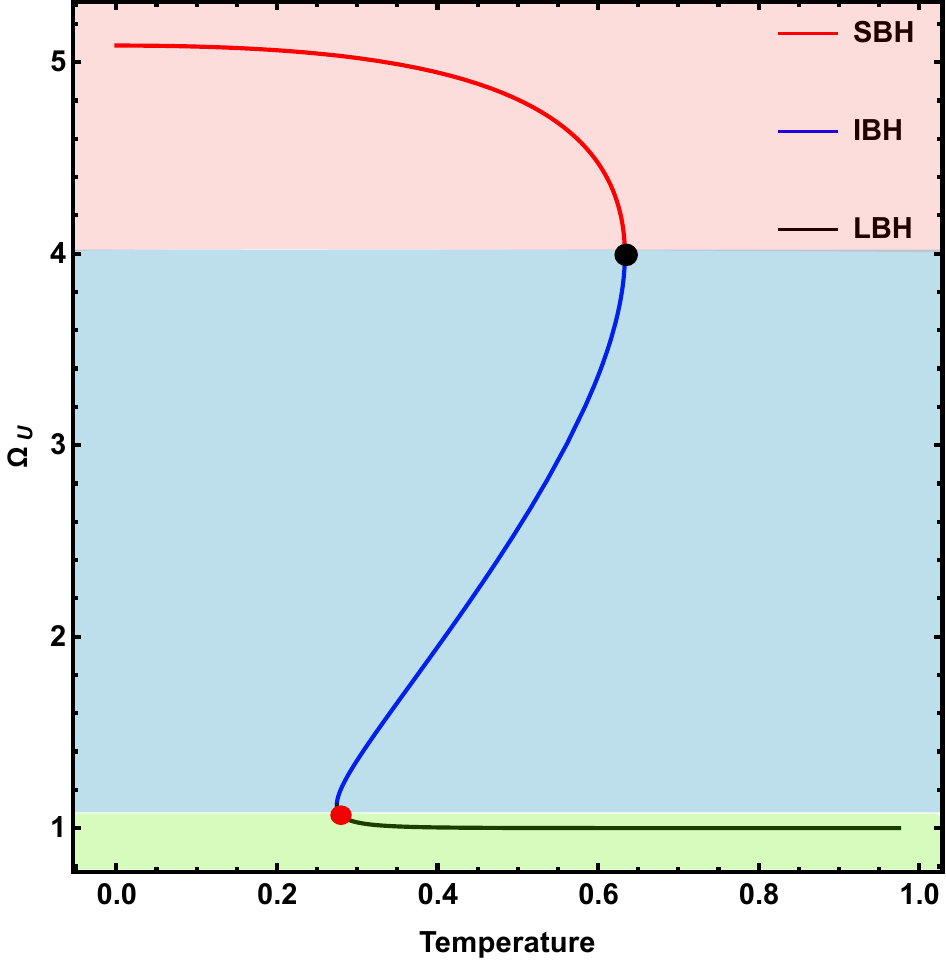}
\includegraphics[width=0.32\linewidth]{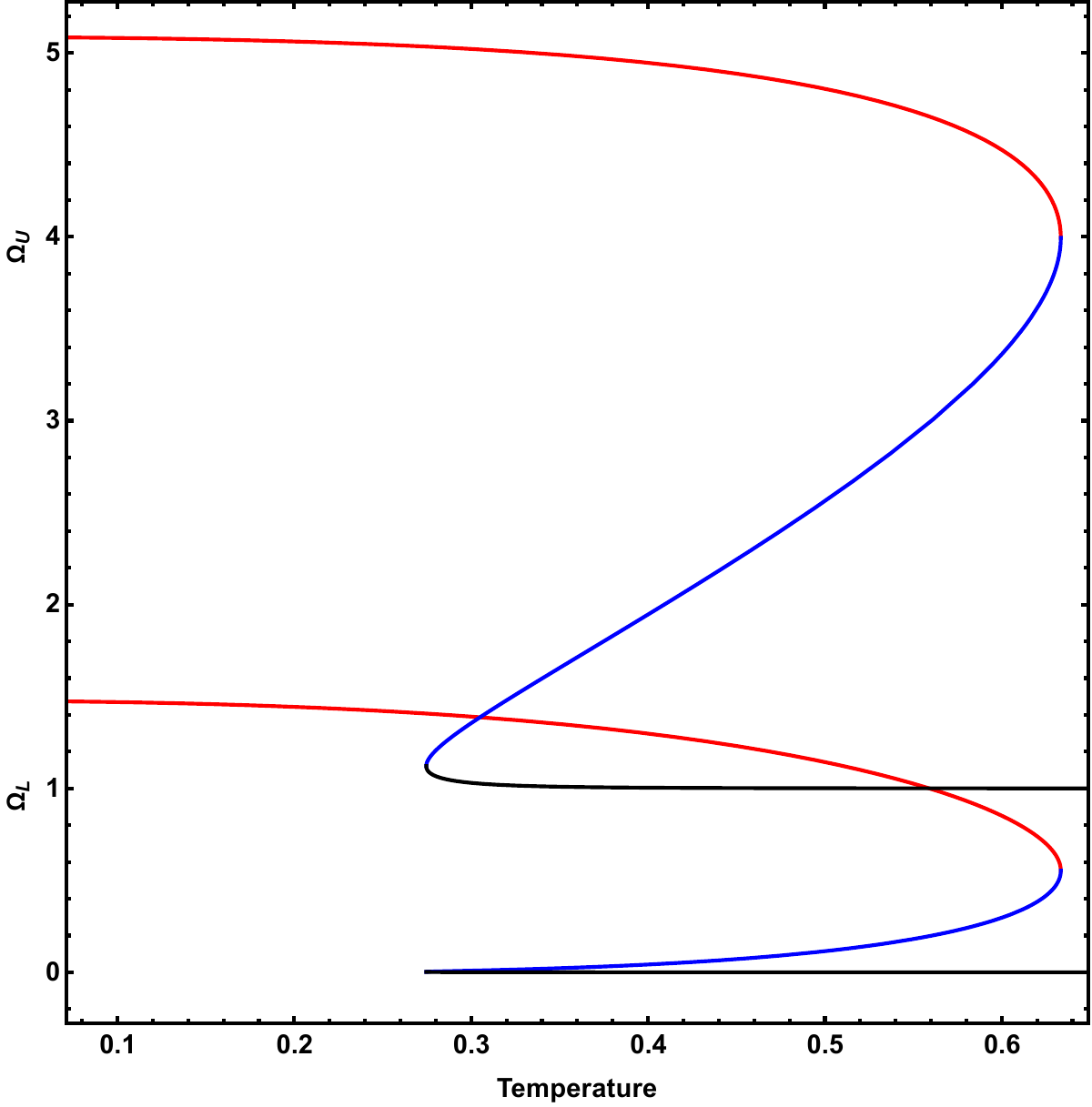}
\vspace{-0.3cm}
\caption{ Behavior of the quasi-periodic oscillation (QPO) frequencies of both upper $\nu_U$
  and lower $\nu_L$ for RN AdS black holes as functions of temperature, across different black hole branches: small (SBH), intermediate(IBH) and large (LBH).   Here, we have considered $Q = 0.05$.Both axes are expressed in natural units.
}
\label{fig2}
\end{figure*}
These expressions characterize the radial and vertical oscillatory response of the particle to small disturbances from the circular orbit in the equatorial plane.  
To convert these frequencies into physical frequencies in units of Hertz (Hz), we employ the following relation:
\begin{align}
    \nu_i = \frac{c^3}{2\pi G M} \cdot \Omega_i
\end{align}

Next we calculate the temperature of the RN black hole using the following formula :
\begin{equation}
T=\frac{-Q^2+3 r_+^4+r_+^2}{4 \pi  r_+^3}
\end{equation}
it is evident that at $r_+=\frac{\sqrt{\sqrt{12 Q^2+1}-1}}{\sqrt{6}}$, temperature goes to zero. Hence $r_+<\frac{\sqrt{\sqrt{12 Q^2+1}-1}}{\sqrt{6}}$ is the unphysical region.Now we plot the radial and vertical oscillatory response of the particle in a unstable circular null geodesics with different values of $Q$ in Fig.\ref{fig1}.  The distinction between physical area and non physical region is clearly observable in the figure.

\subsection{QPO Models}

In this section, we explore the behavior of twin-peak quasi-periodic oscillations (QPOs) in the background of RN AdS black holes. The upper (\(\nu_U\)) and lower (\(\nu_L\)) QPO frequencies are modeled as functions of the radial coordinate and black hole parameters, guided by several well-established theoretical frameworks \cite{51}.

The QPO models considered in our analysis are summarized as follows \cite{51}:
\begin{itemize}
    \item \textbf{Relativistic Precession (RP) model:} In this framework, the upper and lower frequencies correspond to the azimuthal and periastron precession frequencies respectively, i.e., \(\nu_U = \nu_\phi\), \(\nu_L = \nu_\phi - \nu_r\).
    
    \item \textbf{Epicyclic Resonance (ER) models:} These models describe resonant interactions within a geometrically thick disk. The different variants are:
    \begin{itemize}
        \item \textbf{ER2:} \(\nu_U = 2\nu_\theta - \nu_r\), \(\nu_L = \nu_r\),
        \item \textbf{ER3:} \(\nu_U = \nu_\theta + \nu_r\), \(\nu_L = \nu_\theta\),
        \item \textbf{ER4:} \(\nu_U = \nu_\theta + \nu_r\), \(\nu_L = \nu_\theta - \nu_r\).
    \end{itemize}
    
    \item \textbf{Warped Disk (WD) model:} This model, based on motion in a thin accretion disk, defines the frequencies as \(\nu_U = 2\nu_\phi - \nu_r\) and \(\nu_L = 2(\nu_\phi - \nu_r)\).
\end{itemize}

The primary goal of this study is to determine the behavior of the upper and lower QPO frequencies across these different models and examine how they respond to changes in black hole's thermodynamic phases. By comparing the evolution of \(\nu_U\) and \(\nu_L\) across different black hole branches, we aim to identify possible imprints of phase transitions in the QPO spectrum. In doing so, we assess whether these oscillation models can provide insight into the thermodynamic stability and structure of the RN black hole phases.

\subsection{Signatures of Phase transition}
We begin our analysis with the Relativistic Precession (RP) model, where the upper and lower quasi-periodic oscillation (QPO) frequencies are defined as:
\[
\nu_U = \nu_\phi, \qquad \nu_L = \nu_\phi - \nu_r.
\]
By plotting these frequencies against the temperature for subcritical charge values (e.g., $Q = 0.05$), a clear separation of black hole phases becomes evident within the frequency domain.

For $Q = 0.05$, the phase structure can be delineated as follows:  
the small black hole (SBH) branch exists from $(r_+, T) = (0.0498149, 0)$ to $(0.08761, 0.633382)$,  
the intermediate black hole (IBH) branch spans from $(0.08761, 0.633382)$ to $(0.57066, 0.274613)$,  
and the large black hole (LBH) branch extends from $(0.57066, 0.274613)$ towards the asymptotic limit $(r_+, T) \rightarrow (\infty, \infty)$.

As illustrated in Fig.~\ref{fig2}, the solid red curve represents the SBH branch. In this regime, both $\nu_U$ and $\nu_L$ decrease with rising temperature, indicating that the frequency of a particle orbiting the black hole reduces as the system heats up. 
In contrast, along the IBH branch(solid blue curve), both frequencies increase with temperature, suggesting an unstable regime where the particle experiences stronger oscillations.   Finally, in the LBH branch (solid black curve) the frequencies begin to decrease once more and eventually saturate.  When the black hole charge exceeds its critical value, the system exhibits a single thermodynamically stable phase across the entire temperature domain. This is reflected in the frequency behavior as well: for $Q$ above the critical threshold, both $\nu_U$ and $\nu_L$ decrease monotonically with temperature and asymptotically approach constant values. 

It is evident that the QPO behavior of a test particle can serve as an effective probe of black hole phase transitions and their stability. The connection between dynamical frequencies and thermodynamic stability suggests a compelling observational signature for identifying black hole phases. However, further observational data will be crucial to solidify this correspondence. 

\subsection{Massive particle(timelike geodesic)}

\begin{figure*}[!t]
\centering
\includegraphics[width=0.45\linewidth]{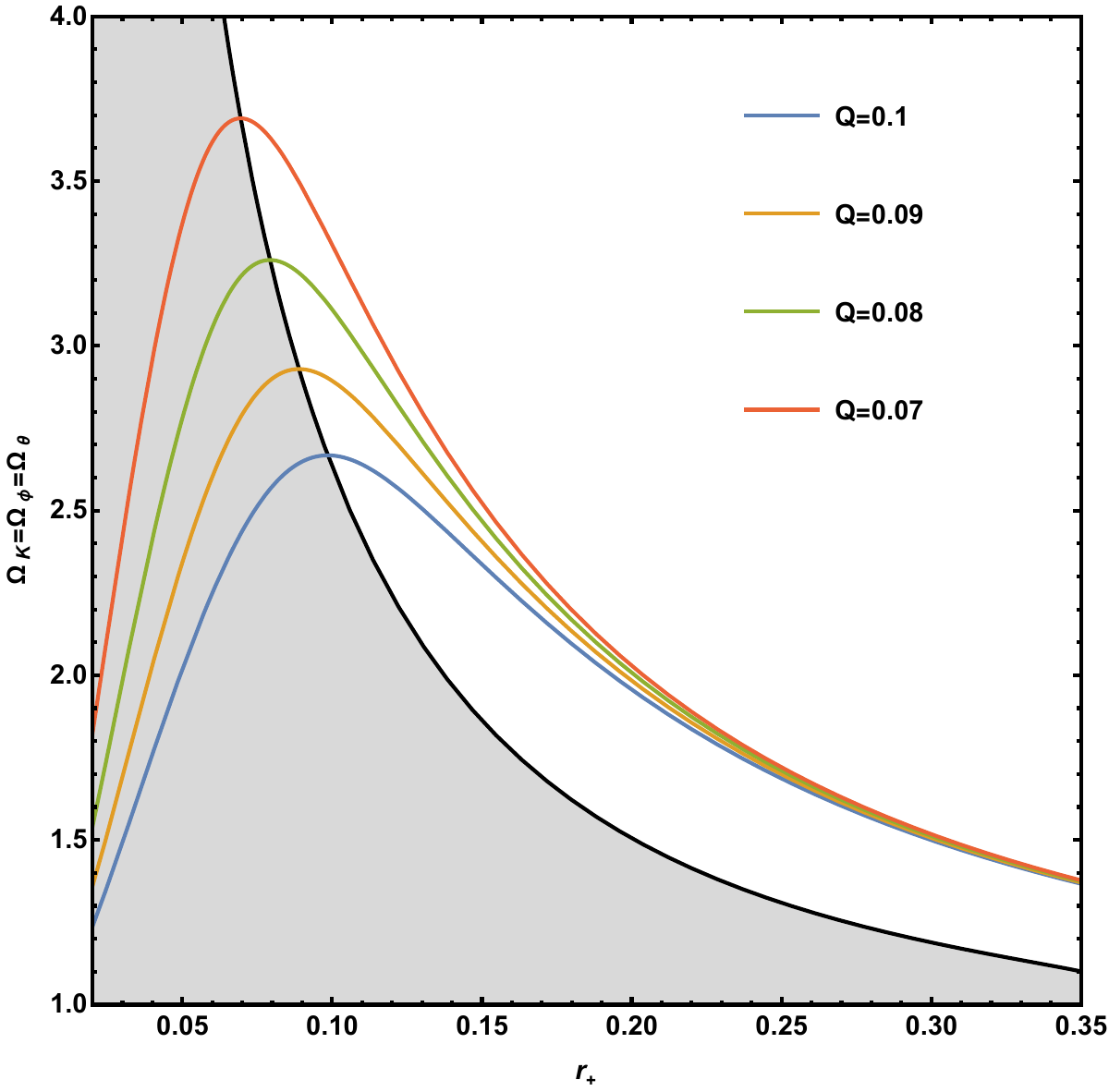}  
\hspace{0.5cm}
\includegraphics[width=0.45\linewidth]{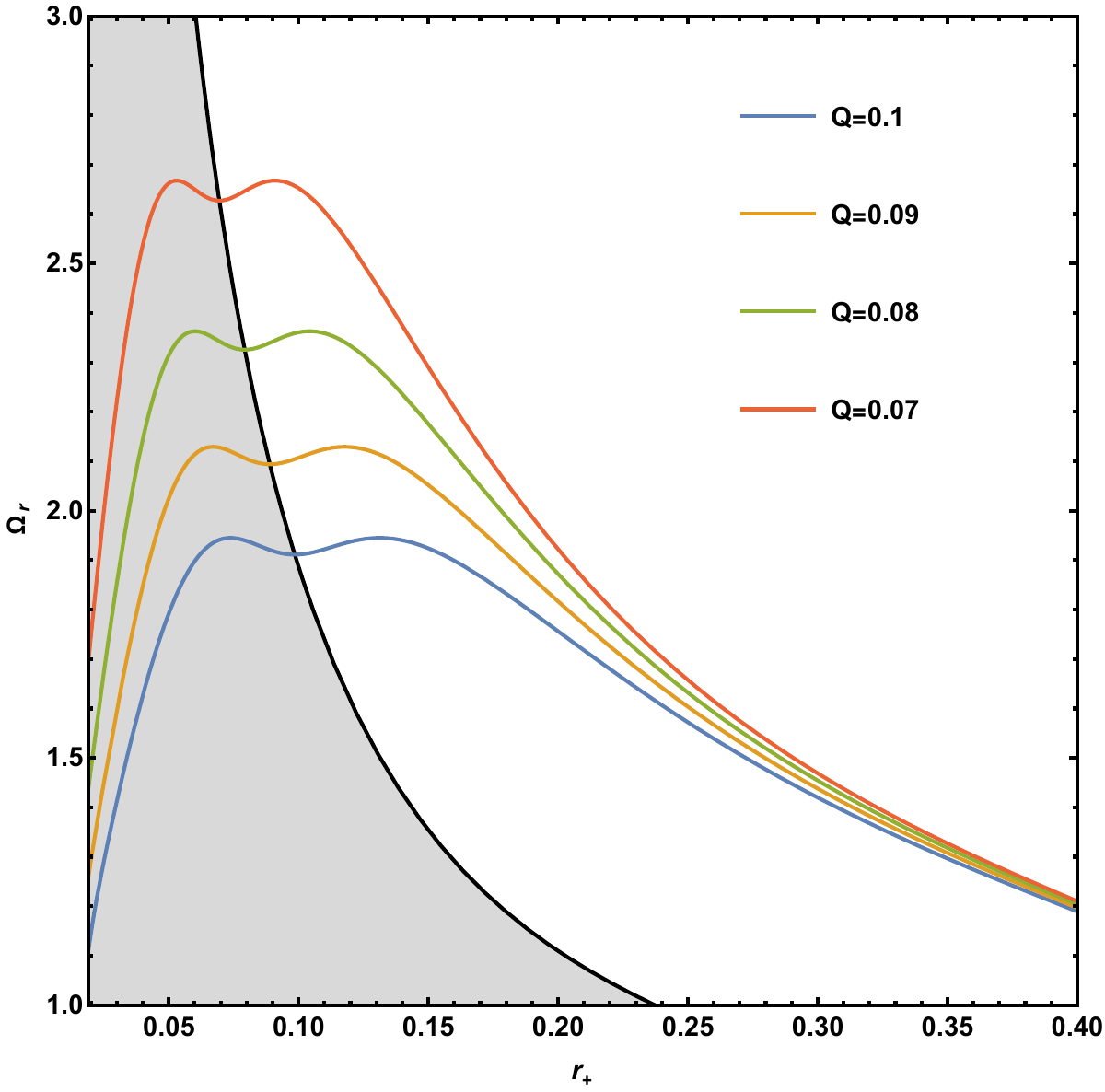}
\vspace{-0.3cm}
\caption{Radial and vertical oscillatory response of the particle in a unstable circular timelike geodesics with different values of $Q$.We have considered L=20.  Gray area represents the non-physical region where temperature is negative.Both axes are expressed in natural units.
}
\label{fig3}
\end{figure*}
In this section, we perform similar analysis for massive particle where we set  $\delta_1=0$ in Eq.\eqref{Eq:m}
The expressions for E and L is obtained to be :
\begin{equation}
E^2=\frac{2 f\left(r_0\right){}^2}{2 f\left(r_0\right)-r_0 f'\left(r_0\right)},
\end{equation}
and
\begin{equation}
L^2=\frac{r_0^3 f'\left(r_0\right)}{2 f\left(r_0\right)-r_0 f'\left(r_0\right)}.
\end{equation}
Next, using the expression of L,  the relation between $r_0$ and $r_+$ is obtained.Here the $r_0$ depends upon the angular momentum $L$ unlike massless particle case.  Using the obtained relation between event horizon radius and radius of the unstable circular orbit we obtained the expressions for radial and vertical frequencies as shown in the previous section.In Fig.\ref{fig3}, we plot the frequencies as a function of event horizon radius $r_+$. The black line represents the temperature zero condition and the shaded portion is the non physical region.
\begin{figure*}[t!]
\centering
\includegraphics[width=0.32\linewidth]{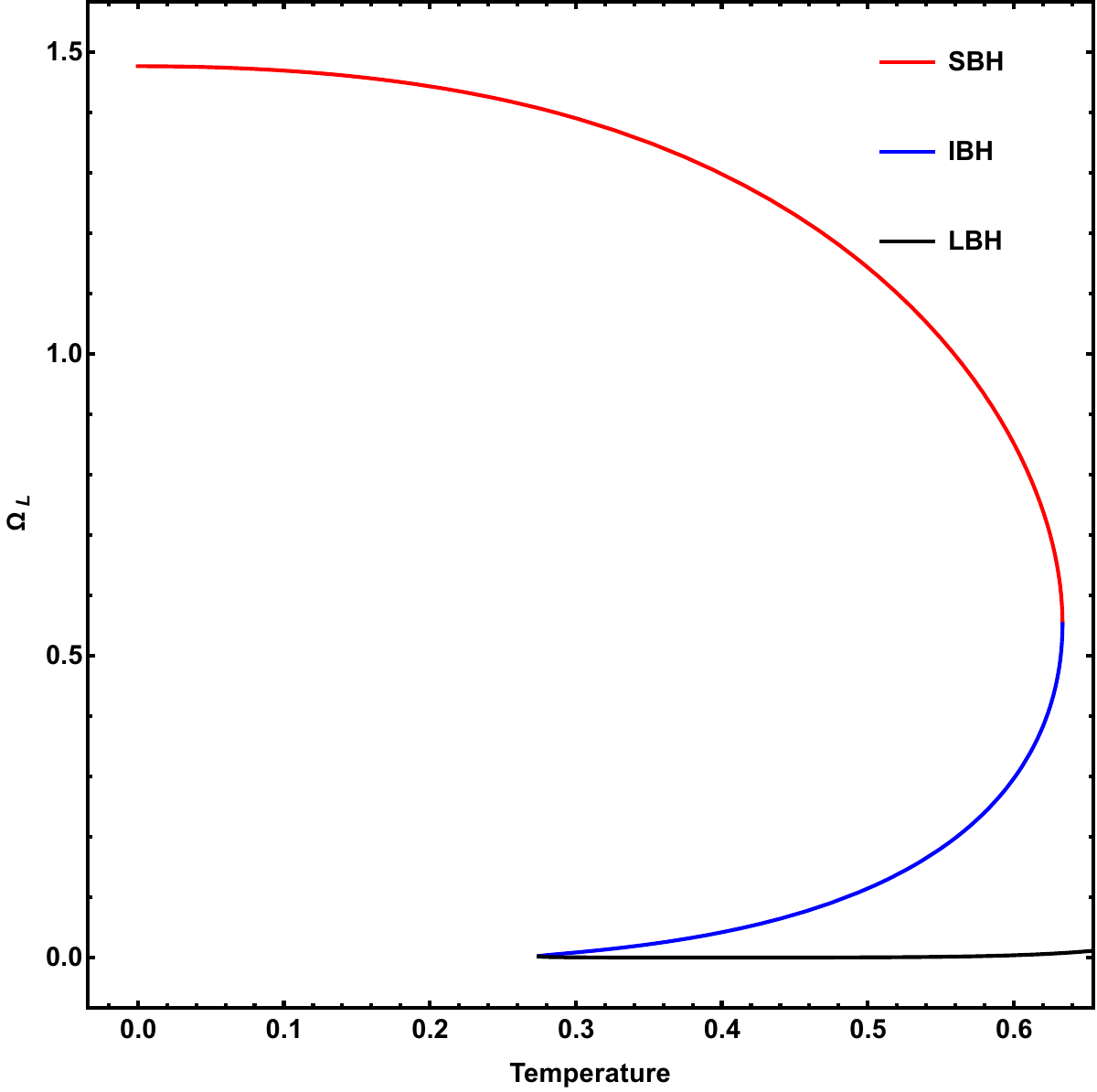}  
\includegraphics[width=0.32\linewidth]{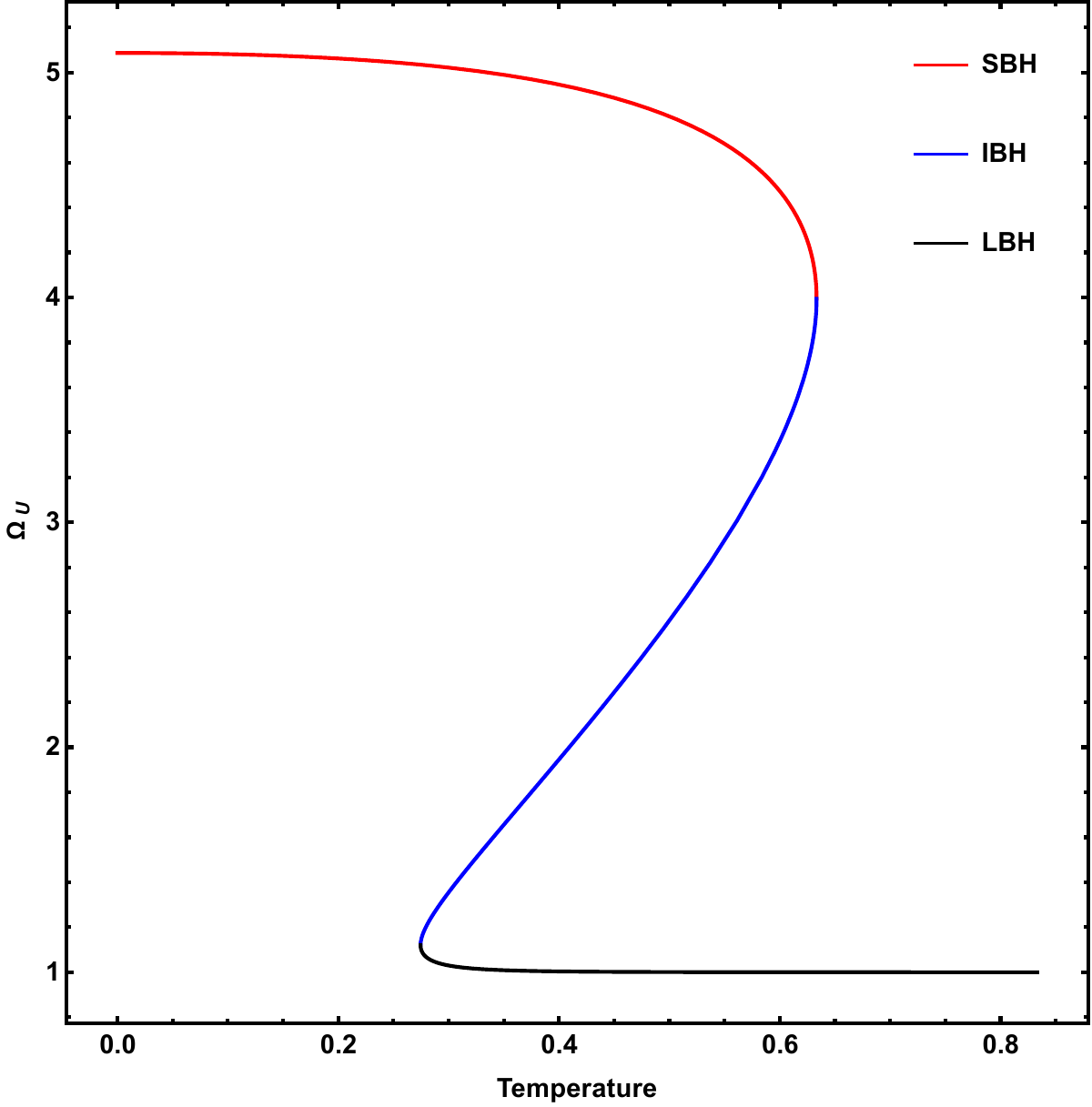}
\includegraphics[width=0.32\linewidth]{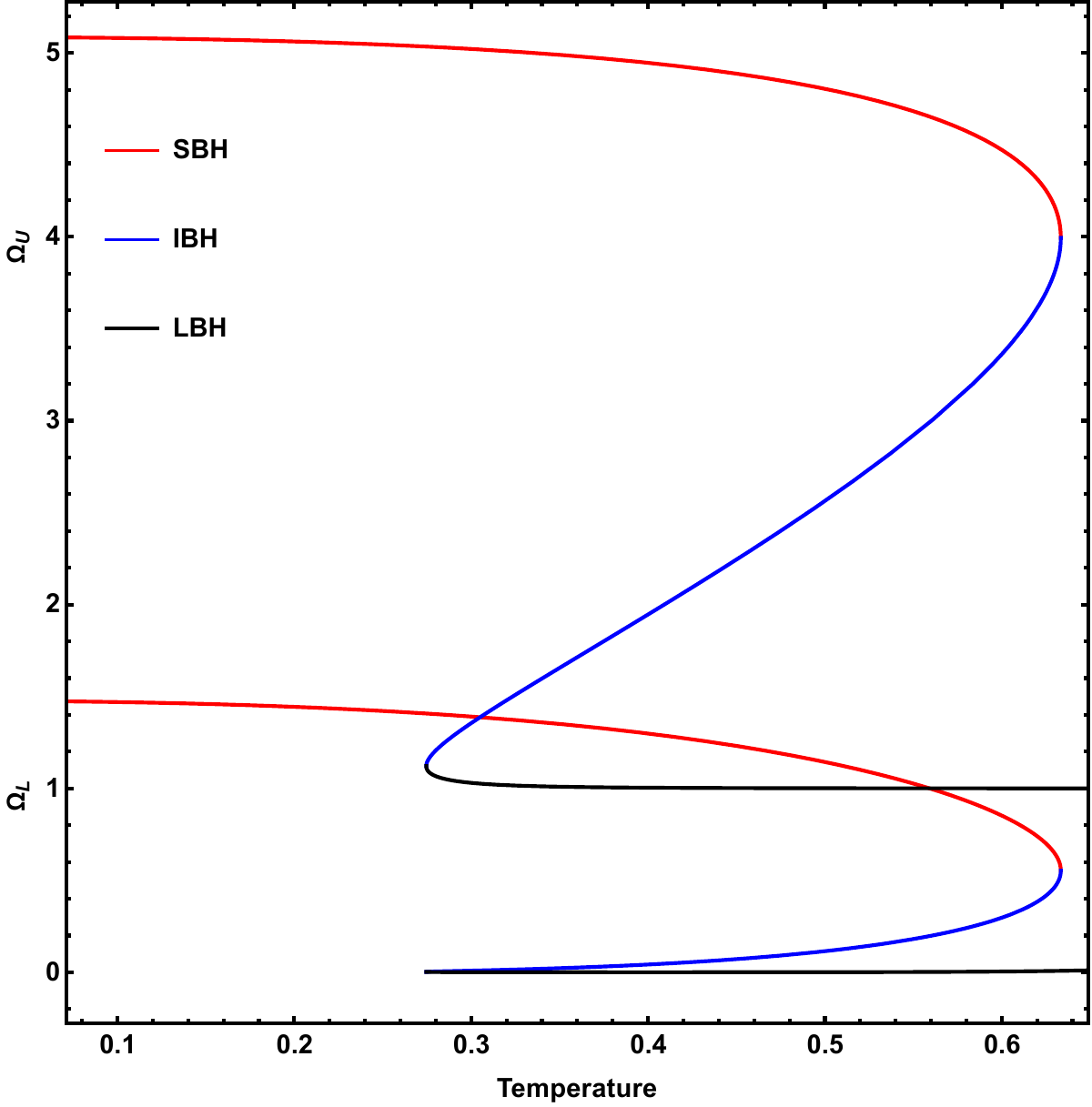}
\vspace{-0.3cm}
\caption{ Behavior of the quasi-periodic oscillation (QPO) frequencies of both upper $\nu_U$
  and lower $\nu_L$ for RN black holes as functions of temperature, across different black hole branches: small (SBH), intermediate(IBH) and large (LBH).   Here, we have considered $Q = 0.05$.Both axes are expressed in natural units.
}
\label{fig4}
\end{figure*}

Now we compute the upper and lower QPO frequencies within the Relativistic Precession (RP) model for massive particles, where the identification of frequencies remains the same:
\[
\nu_U = \nu_\phi, \qquad \nu_L = \nu_\phi - \nu_r.
\]

As in the massless case, we analyze the behavior of these frequencies as functions of the black hole temperature for a subcritical charge value, e.g., \(Q = 0.05\). The frequency plots once again reveal the underlying phase structure of the black hole: the small black hole (SBH) branch spans from $(r_+, T) = (0.0498149, 0)$ to $(0.08761, 0.633382)$, followed by the intermediate black hole (IBH) branch from $(0.08761, 0.633382)$ to $(0.57066, 0.274613)$, and finally the large black hole (LBH) branch extends from $(0.57066, 0.274613)$ toward the asymptotic regime.

In the SBH  and LBH region, both \(\nu_U\) and \(\nu_L\) decrease with increasing temperature.  In the IBH branch, the frequencies increase with temperature.  For charge values above the critical threshold (\(Q > Q_c\)), the temperature-frequency relation is smooth and monotonic, with no indication of a phase transition. In this regime, both frequencies evolve predictably: \(\nu_U\) decreases gradually with temperature, while \(\nu_L\) remains positive and asymptotically approaches zero from above.  In summary,  the overall qualitative behavior of QPO frequencies for massive particles is consistent with the massless case.\\

It is important to note that the QPO frequencies observed are believed to be influenced by the temperature and dynamics of the accretion disk, which can reach values of \(10^6\)–\(10^7\) K and is accessible via X-ray spectral observations. In contrast, the Hawking temperature of a black hole,  is an extremely small quantity and has no direct observational signature as of now.  However, changes in the Hawking temperature are associated with changes in the geometry of the black hole spacetime, particularly near the event horizon. Since QPO frequencies are sensitive to the spacetime structure in the inner accretion region, these geometric changes can influence their behavior.  We do not suggest a direct physical connection between the two,  but rather use the temperature as a thermodynamic indicator of the black hole's geometric phase.  By plotting QPO frequencies directly against Hawking temperature, we aim to isolate the influence of the black hole geometry on the QPO behavior. Although the exact physical mechanism connecting the two remains complex and not fully understood, the way QPO frequencies change across different thermodynamic phases reveals meaningful patterns. \\

We have also performed our analysis using the ER2, ER3, ER4 models as well as the Warped Disk (WD) model for both mass and massless particle case.  All these QPO models exhibit consistent qualitative behavior and successfully capture the features of black hole phase transitions and the associated stability structure. For the sake of brevity, we omit the detailed results here while these analysis is shown for Kerr black hole as it will be more physically meaningful.  It is worth noting, however, that while the overall trends remain coherent across models, the actual frequency values may vary.  Across all five models, the behavior of the upper and lower QPO frequencies as functions of temperature remains qualitatively similar: in the SBH phase, frequencies decrease with temperature, indicating stable motion; in the IBH phase, the frequencies rise, signaling instability; and in the LBH phase, they gradually level off and approach constant values in the asymptotic limit.

\section{ Kerr black hole}
\begin{figure*}[t!]
\centering
\includegraphics[width=0.32\linewidth]{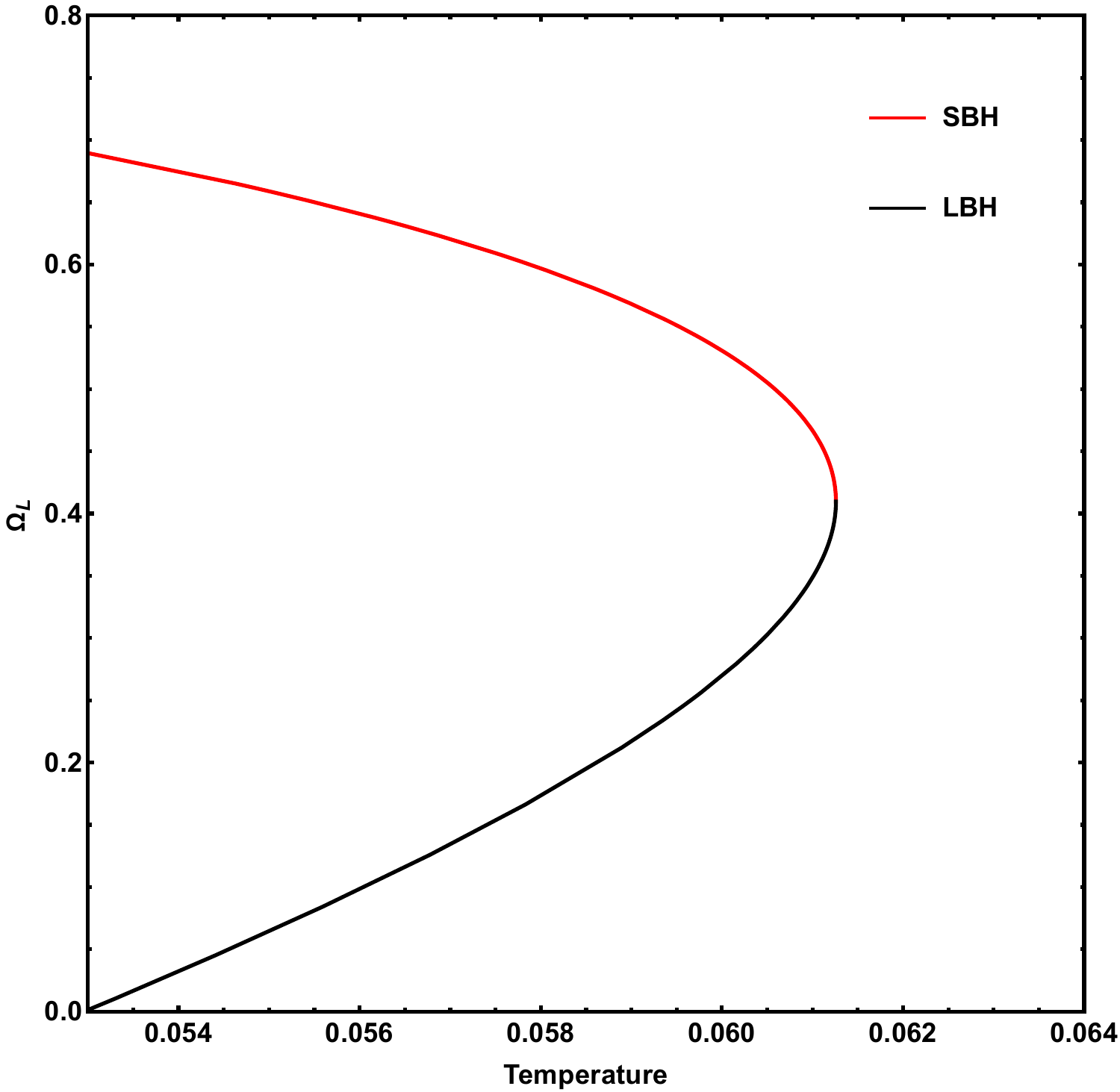}  
\includegraphics[width=0.32\linewidth]{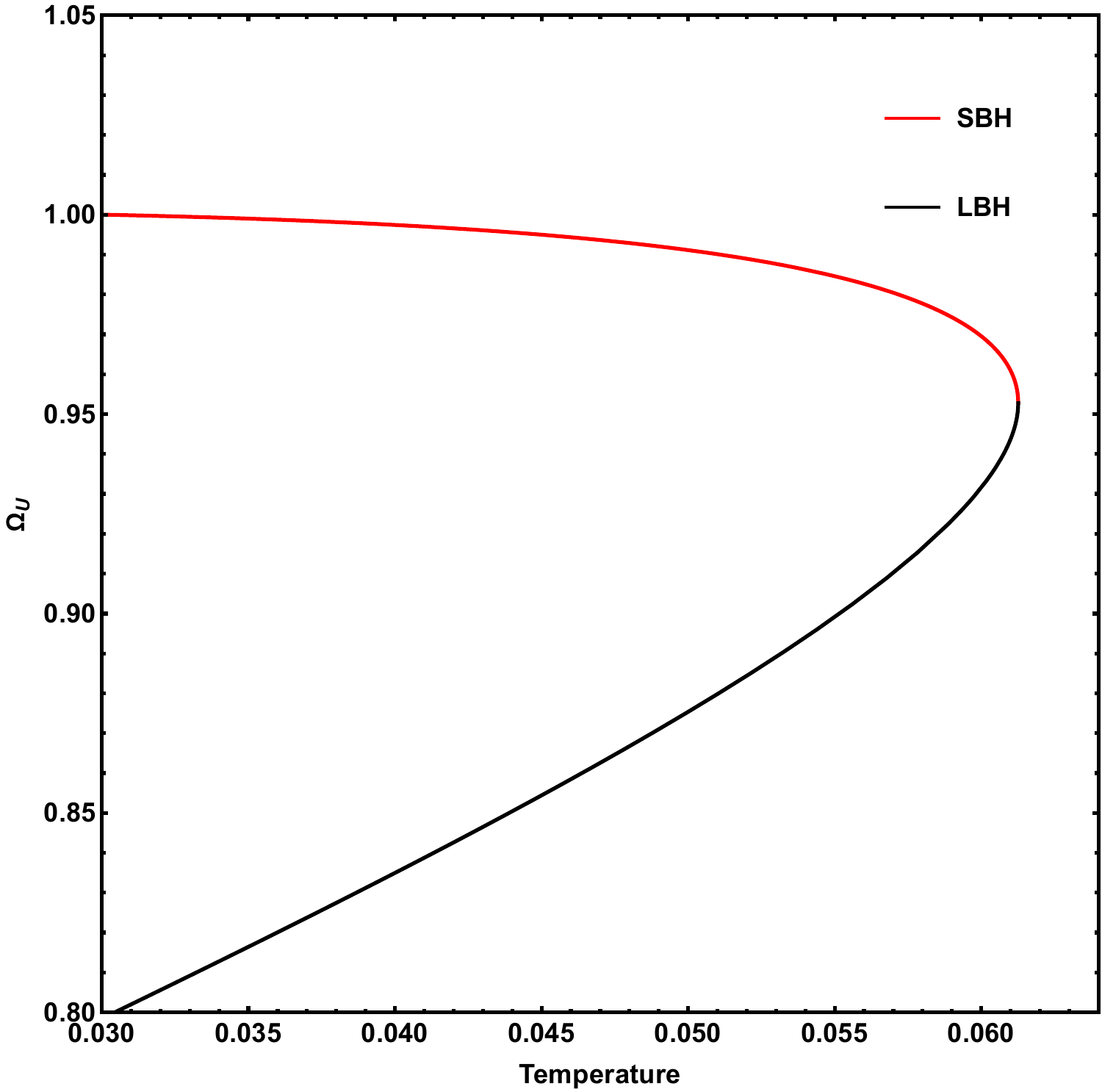}
\includegraphics[width=0.32\linewidth]{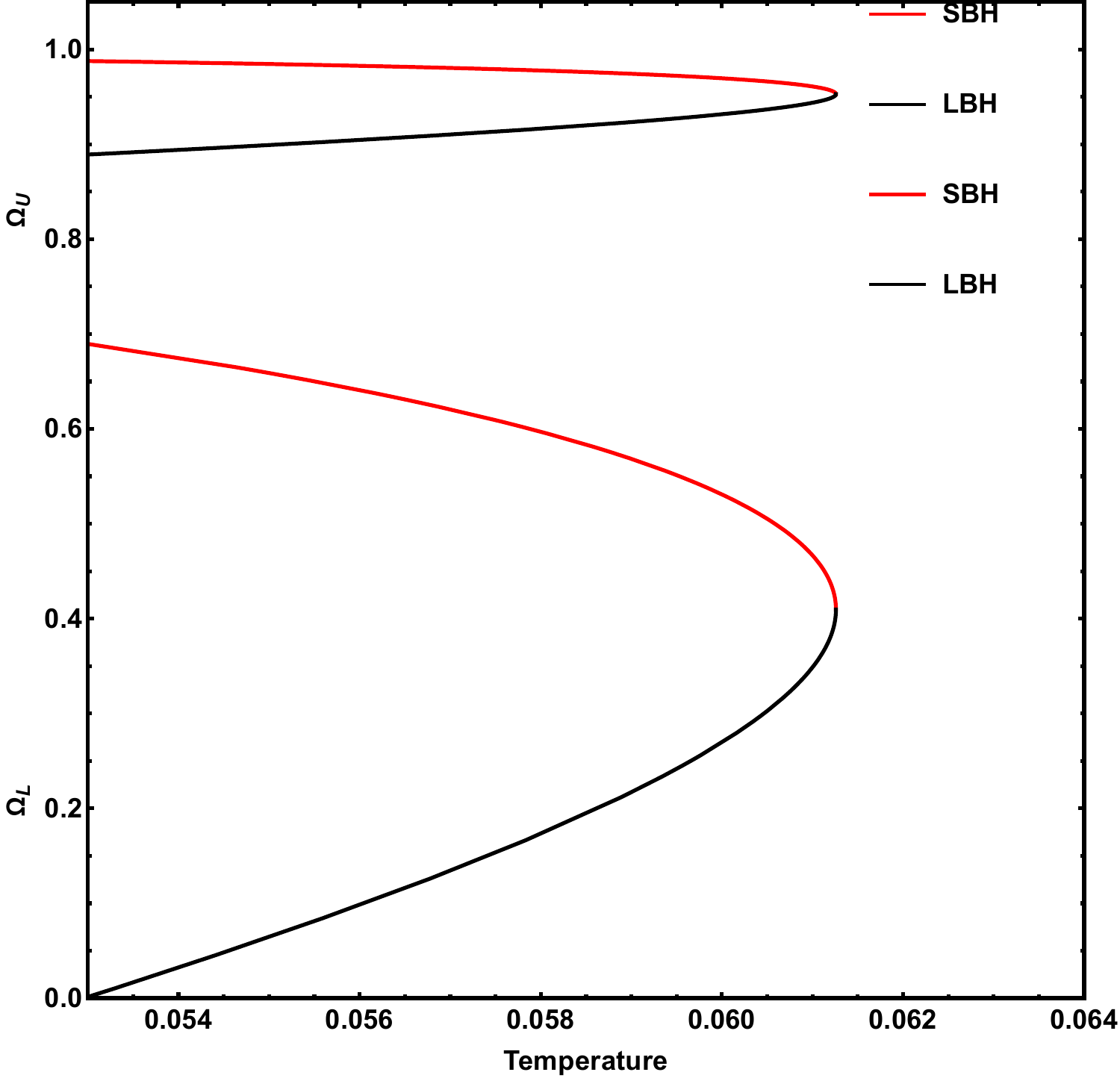}
\vspace{-0.3cm}
\caption{ Behavior of the quasi-periodic oscillation (QPO) frequencies of both upper $\nu_U$
  and lower $\nu_L$ for Kerr black holes as functions of temperature, across different black hole branches in RP model : small (SBH) and large (LBH).   Here, we have considered $a = 0.5$.Both axes are expressed in natural units.
}
\label{fig5}
\end{figure*}
To ensure astrophysical relevance, we adopt the Kerr black hole geometry as the background spacetime.  We have considered the black holes to be rotating and asymptotically flat, making them more consistent with the nature of astrophysical black holes observed in X-ray binaries and active galactic nuclei. Rotation is a crucial feature in accreting black hole systems, as the presence of an accretion disk inherently implies angular momentum.  In this case we only consider the massive particle case, as it will be more physically meaningful.
\subsection{Thermodynamics of Kerr Black hole}
The mass of the black hole is 
\begin{equation}
M=\frac{a^2+r_+^2}{2 r_+}
\end{equation}

The temperature is calculated as 
\begin{equation}
T=\frac{r_+^2-a^2}{4 \pi  r_+^3}
\end{equation}
\begin{figure}
\includegraphics[height=6cm,width=8cm]{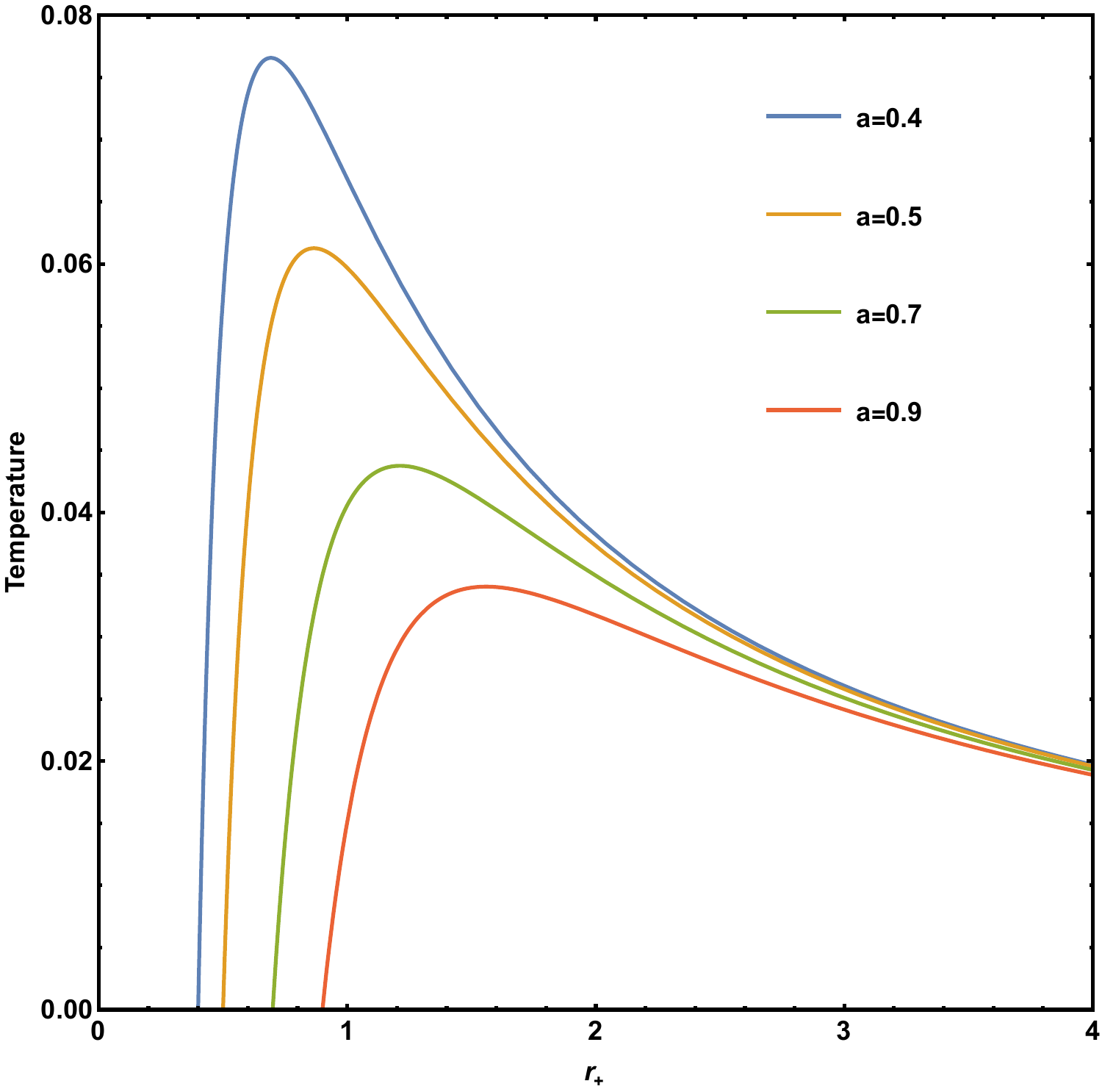}
\caption{Temperature of the Kerr black hole with respect to event horizon radius.Both axes are expressed in natural units.
}
\label{fig6}
\end{figure}
In the temperature plot, we observe two black hole branch. As the rotation $a$ value increases, the temperature value decreases. The SBH branch is found to be thermodynamically stable while the LBH branch is found to be unstable.  We next study the QPO profiles. 

The effective potential in case of a massive particle orbiting around a Kerr black hole is 
\begin{multline}
V_{eff}=\\
\frac{-\left(\text{E} \left(a^2+r^2\right)-a L \Xi \right)^2+\Delta (r) (a \text{E}-L \Xi )^2+r^2 (-\Delta (r))}{r^4}
\end{multline}
Accordingly,  the squared radial and vertical frequencies in case of Kerr black holes can be calculated as 
\begin{equation}
\Omega_r^2=\frac{M r \left(r (r-6 M)-3 a^2\right)+8 a (M r)^{3/2}}{r^2 \left(a \sqrt{M r}+r^2\right)^2}
\end{equation}
\begin{equation}
\Omega_\theta^2=\frac{3 a^2 M r-4 a M r \sqrt{M r}+M r^3}{\left(a r \sqrt{M r}+r^3\right)^2}
\end{equation}
\begin{equation}
\Omega_\phi^2=\frac{M r}{\left(a \sqrt{M r}+r^2\right)^2}
\end{equation}

Next, we revisit the Relativistic Precession (RP) model to compute the upper and lower QPO frequencies as functions of the event horizon radius \( r_+ \). Using these results, we construct a parametric plot between the black hole temperature and the corresponding QPO frequencies. The plot reveals the existence of two distinct black hole branches.
For the small black hole (SBH) branch, we observe that the QPO frequencies decrease with increasing temperature. This monotonic behavior indicates thermodynamic stability, as the system responds smoothly to changes in temperature. In contrast, along the large black hole (LBH) branch, the QPO frequencies exhibit an increasing trend with temperature. This reversal in behavior suggests thermodynamic instability in the LBH regime.

Our analysis reveals that the slope of the QPO frequency–temperature relation, as represented by \(\arctan(\nu_L/T)\), is notably influenced by the black hole's spin parameter \(a\). As the rotation increases, we observe that the slope steepens in specific regions, especially near critical points where frequency behavior changes rapidly. A steeper slope indicates a stronger sensitivity of the  QPO frequency to variations in temperature, suggesting that even small thermodynamic changes in the black hole result in noticeable shifts in QPO characteristics.  Moreover, increasing spin appears to shift the onset of these rapid variations to larger horizon radii, implying a delay in the transition behavior.  \\

The signatures observed in the QPO frequency–temperature relations likely arise from an indirect but interesting connection between black hole thermodynamics and the motion of matter in the surrounding spacetime. When a black hole undergoes a thermodynamic transition such as moving from a small to a large black hole phase it changes the geometry near the event horizon. These geometric changes affect the effective potentials that determine how particles and fluid elements move around the black hole. Since QPO frequencies are directly related to this motion, any shift in the spacetime geometry can influence the observed oscillation frequencies. The black hole's spin plays a key role in this process. Increasing the spin not only changes the temperature and thermodynamic behavior of the black hole, but also alters the shape of the spacetime and the stability of particle orbits. As a result, the response of QPO frequencies to changes in temperature becomes more sensitive in rapidly rotating black holes. \\

\begin{figure*}[t!]
\centerline{
\includegraphics[scale=0.3]{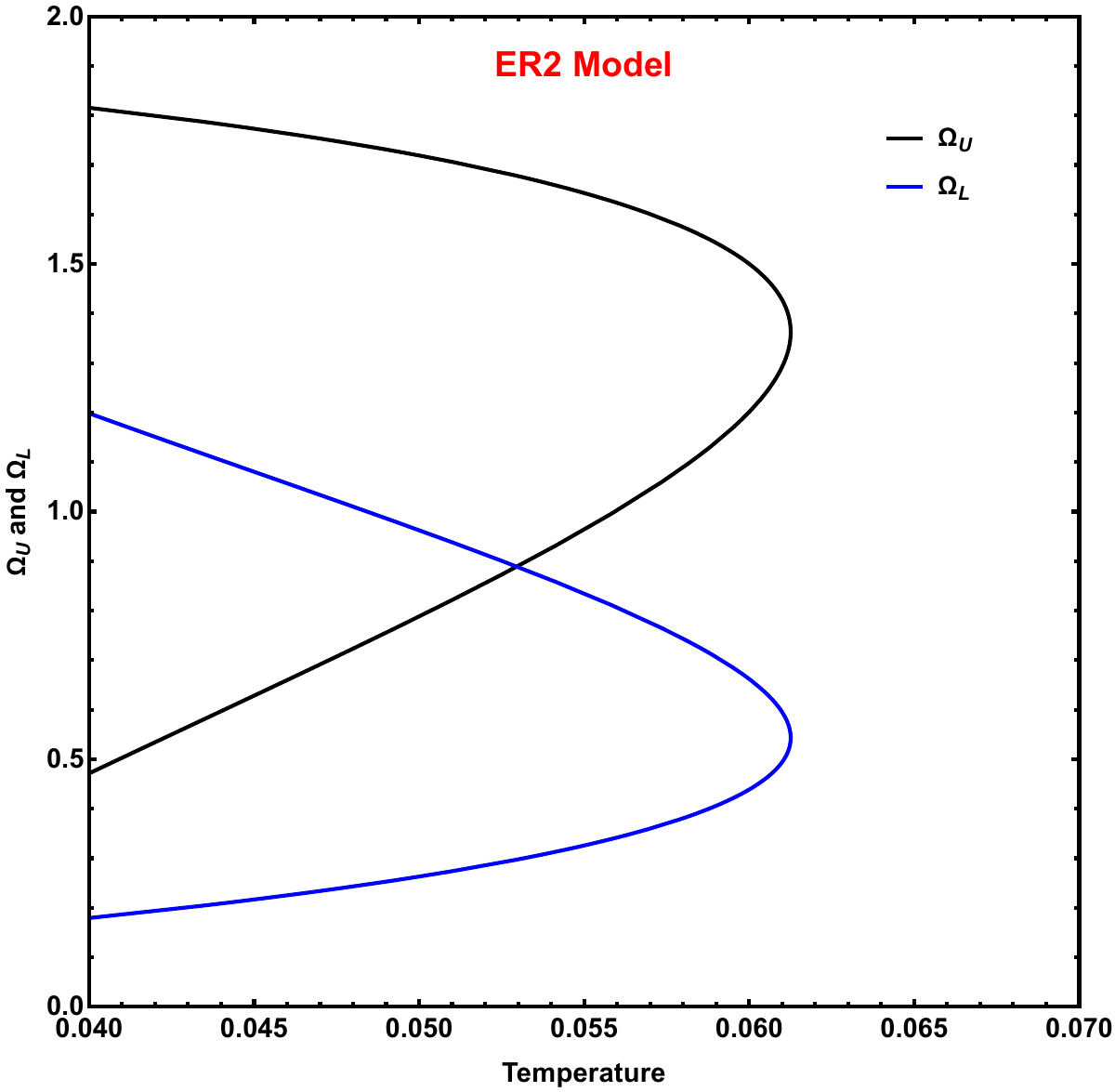} \hspace{20pt} 
\includegraphics[scale=0.3]{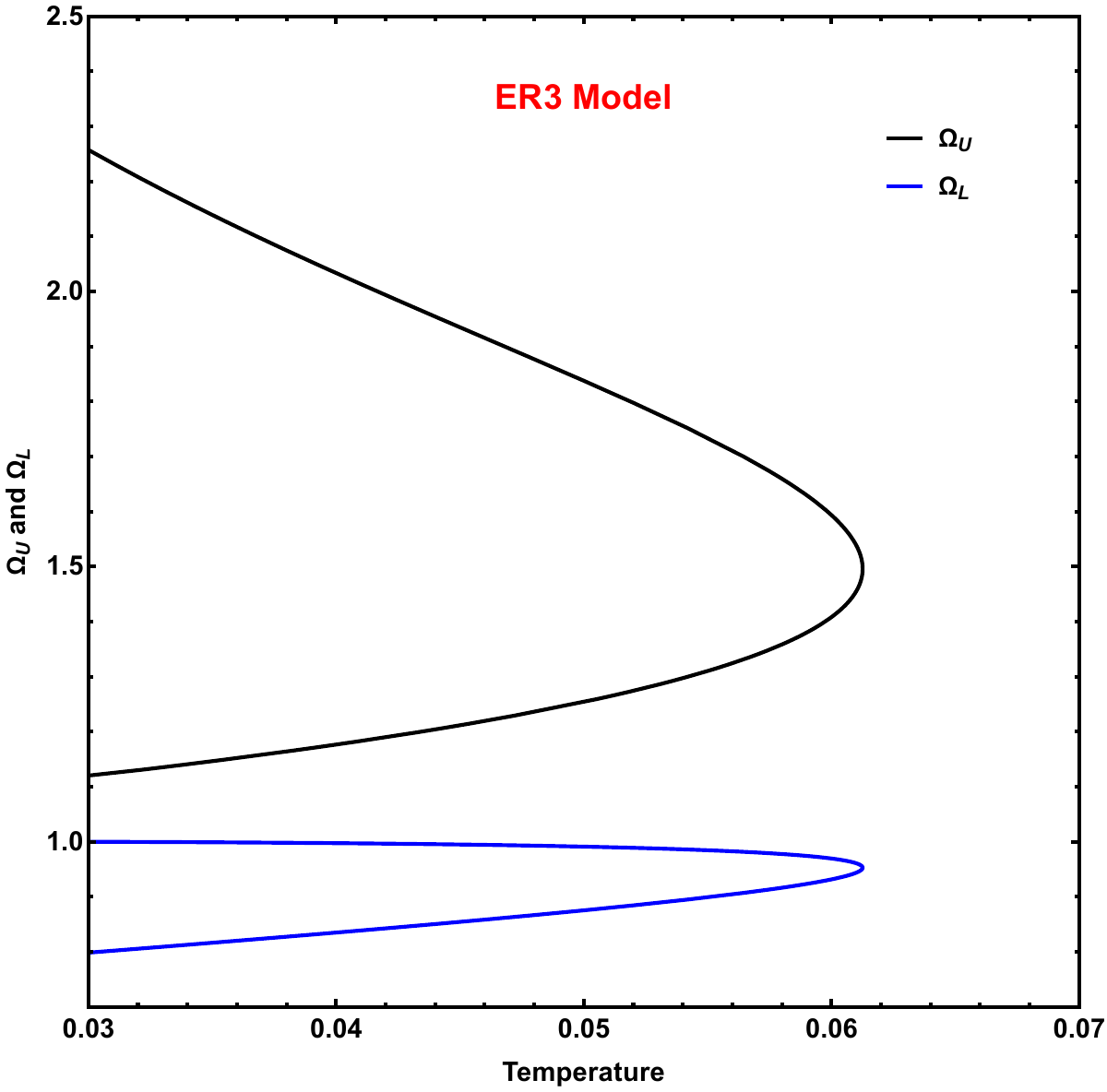}}
\centerline{
\includegraphics[scale=0.3]{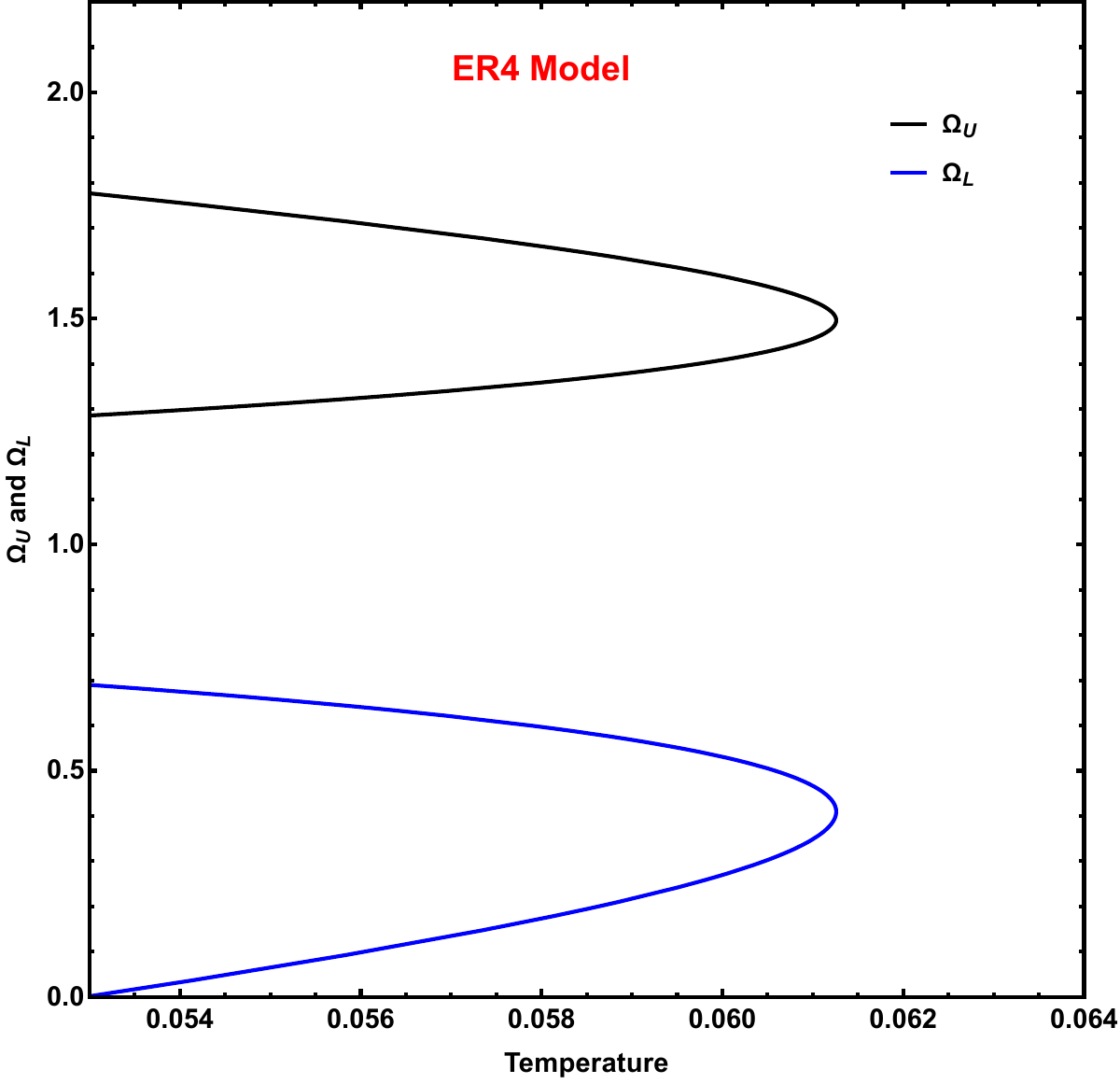}\hspace{20pt}
\includegraphics[scale=0.3]{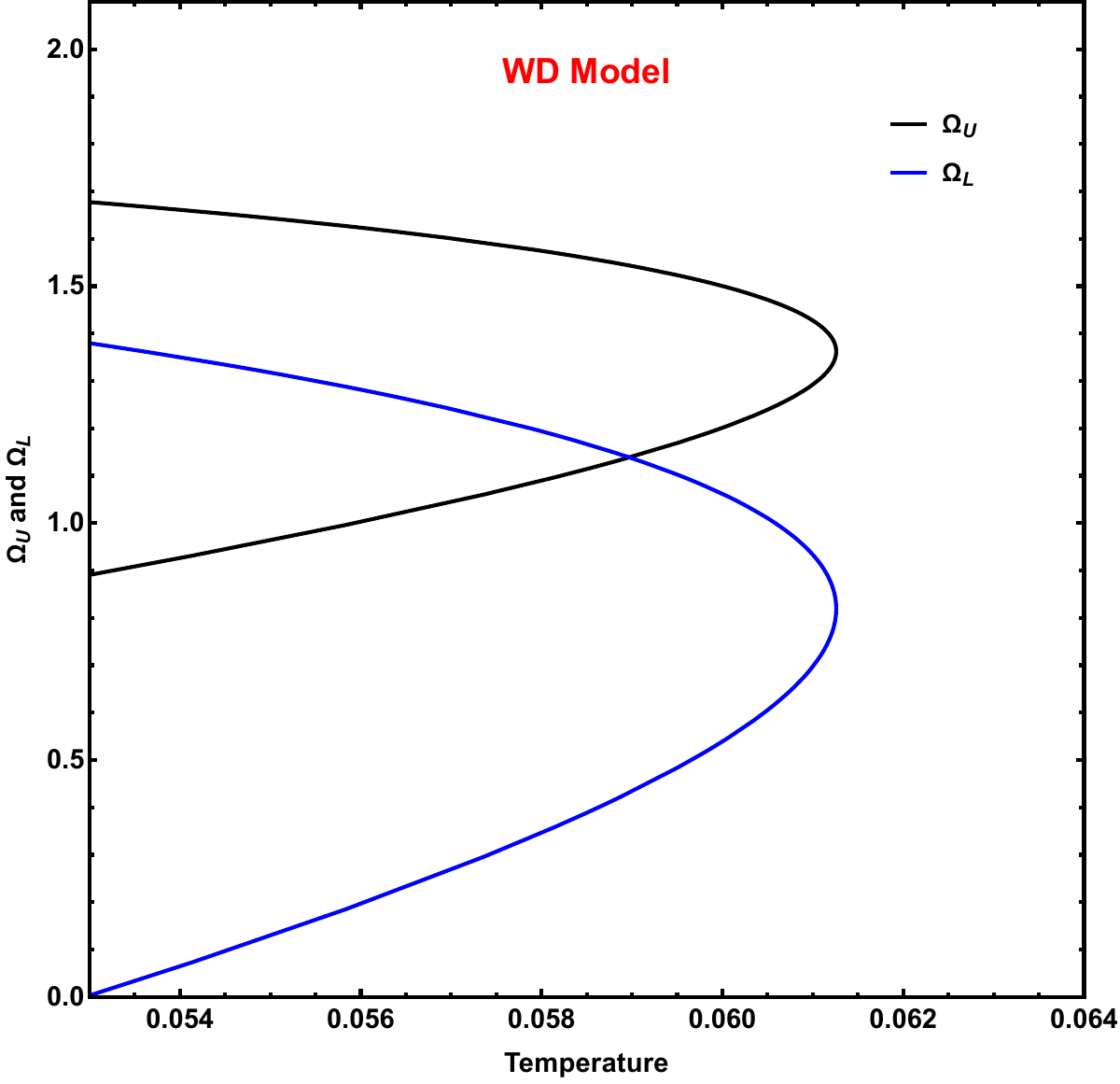}}
\vspace{-0.3cm}
\caption{ Behavior of the quasi-periodic oscillation (QPO) frequencies of both upper $\nu_U$
  and lower $\nu_L$ for Kerr black holes as functions of temperature, across different QPO models.   Here, we have considered $a = 0.5$.Both axes are expressed in natural units.
}
\label{models}
\end{figure*}

We performed our analysis across five distinct QPO models: the Relativistic Precession (RP) model, three variants of the Epicyclic Resonance (ER) model namely ER2, ER3, and ER4 and the Warped Disk (WD) model. For each model, we computed the upper and lower QPO frequencies as functions of the event horizon radius and black hole spin. Despite differences in their construction and frequency relations, all five models exhibit consistent qualitative behavior with respect to black hole thermodynamic structure as shown in Fig.\ref{models} In particular, the characteristic signatures of the small and large black hole branches, as well as the influence of rotation on QPO–temperature response, remain robust across models.  Among these, the ER3 model yields the highest frequency values, as evident from Fig.~8, which aligns with its formulation involving higher-order resonances. \\
.

\section{Comparison with observed HFQPO }

\begin{figure}[H]
\includegraphics[scale=0.3]{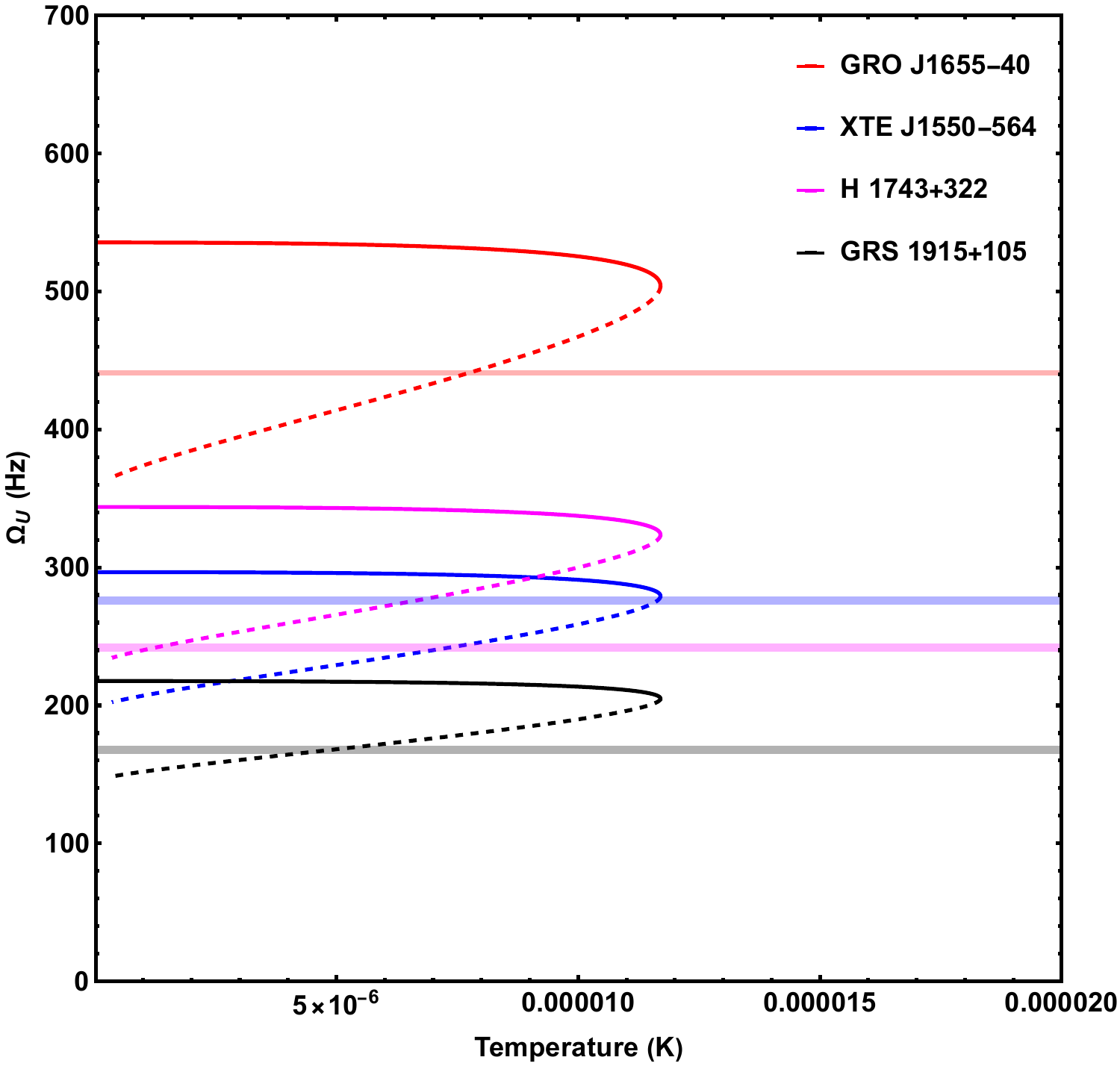} \hspace{20pt} 
\includegraphics[scale=0.3]{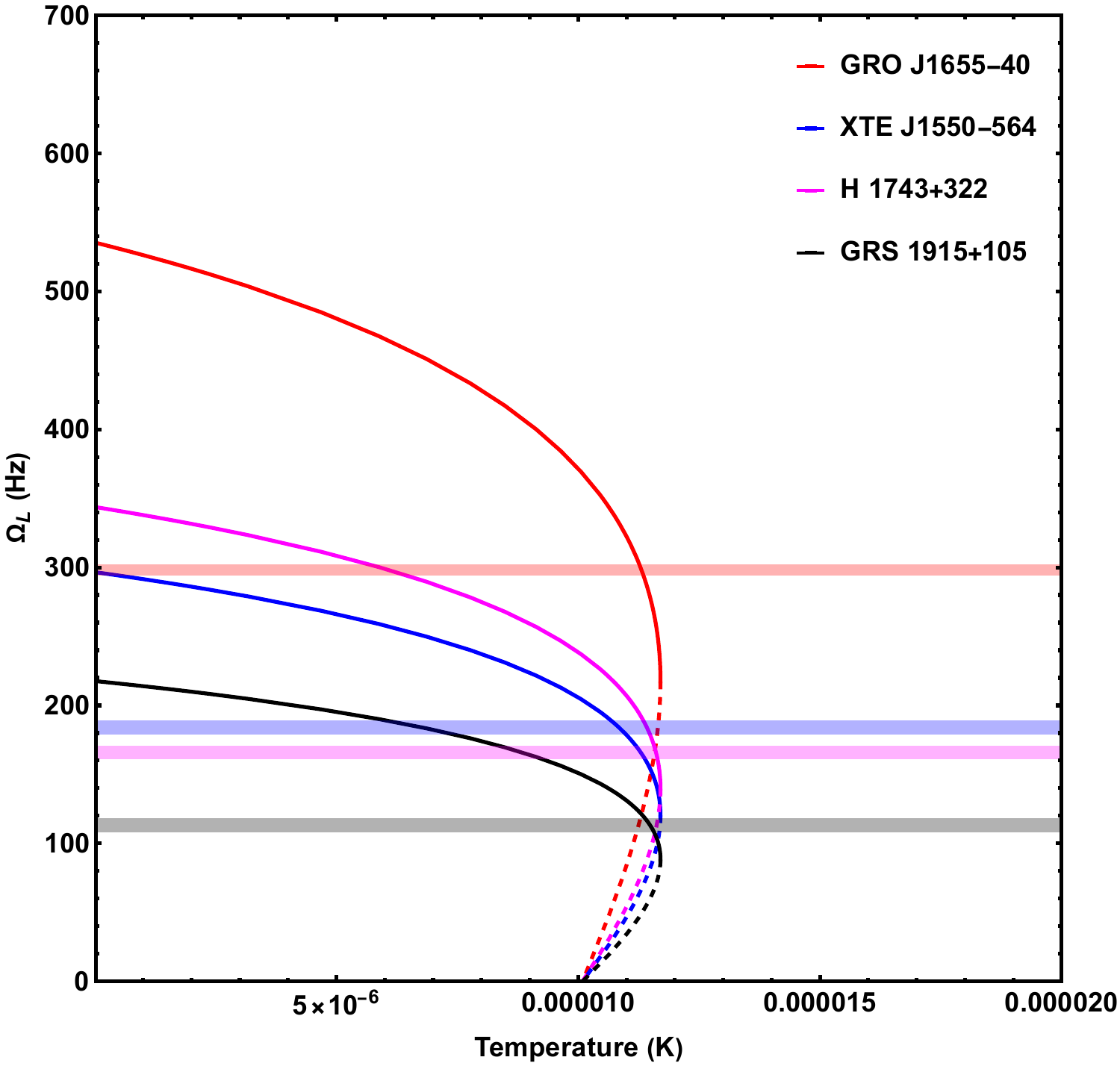}
\caption{ Behavior of the quasi-periodic oscillation (QPO) frequencies of both upper $\nu_U$
  and lower $\nu_L$ for Kerr black holes as functions of temperature.  Here, we have considered $a = 0.5$. The solid line represents the SBH phase while the dotted line represents the LBH phase. }
\label{obs1}
\end{figure}

We now compare our theoretical predictions with the observational data. 
The relevant high-frequency QPO sources, along with their masses and observed 
frequencies, are summarized in Table~\ref{tab1}.  
To make a meaningful comparison with theory, we convert the model predictions 
into physical units using the appropriate scaling relations.  
\begin{equation}
T (Kelvin)= T*\frac{\hbar c}{ k_B } \quad \quad   \nu_i = \frac{c^3}{2\pi G M} \cdot \Omega_i
\end{equation}
{\LARGE
\begin{table*}[htb]
\centering
\begin{tabular}{|c|c|c|c|}
\hline
\textbf{Source} & \textbf{Mass} (in $M_\odot$) & \textbf{Upper Frequency (Hz)} & \textbf{Lower Frequency (Hz)} \\
\hline
GRO J1655$-$40 & $5.4 \pm 0.3$ \cite{t57} & $441 \pm 2$ \cite{t58} & $298 \pm 4$ \cite{t58} \\
\hline
XTE J1550$-$564 & $9.1 \pm 0.61$ \cite{t59} & $276 \pm 3$ & $184 \pm 5$ \\
\hline
GRS 1915$+$105 & $12.4^{+2.0}_{-1.8}$ \cite{t60} & $168 \pm 3$ & $113 \pm 5$ \\
\hline
H 1743$+$322 & $8.0 - 14.07$ \cite{t61,t62,t63} & $242 \pm 3$ & $166 \pm 5$ \\
\hline
\end{tabular}
\caption{Observational QPO data for different Black hole sources with estimated mass  \cite{52}}
\label{tab1}
\end{table*}
}
Throughout this section, we adopt the Kerr black hole spacetime together with the 
relativistic precession (RP) model as our working framework.  
For massless test particles, the resulting behaviours are displayed in 
Fig.~\ref{obs1}.  In the left panel of Fig.~\ref{obs1}, we plot the upper QPO frequency 
as a function of Hawking temperature.  The horizontal coloured bands (red, magenta, blue and black) represent the observed frequencies of GRO~J1655$-$40, H~1743$+$322, XTE~J1550$-$564, and 
GRS~1915$+$105, respectively.   From this analysis, we find that the observed upper HFQPO frequencies for all these sources correspond to the LBH branch.

In contrast, the right panel of Fig.~\ref{obs1} shows the behaviour of the 
lower QPO frequency.  
Here, the same set of observational frequency bands intersect exclusively with 
the SBH branch.  
Thus, the upper and lower peaks point to  different thermodynamic phases at the same time.
This apparent inconsistency indicates that 
HFQPOs are highly sensitive to several astrophysical factors such as accretion 
disk structure, disk magnetosphere coupling, and hydrodynamic or 
magnetohydrodynamic processes.
Consequently, the observed HFQPO values are influenced not only by the 
black hole  geometry but also by the detailed disk physics.Therefore, it is natural that the upper and lower HFQPOs do not predict the same black hole phase, and one cannot infer the thermodynamic state of the black hole solely from a single QPO peak.  
Although only qualitative in nature, the analysis based on the upper HFQPO allows us to infer a tentative range for the black hole radius or phase.  In the framework of the RP model, the upper and lower highfrequency QPOs are identified with different fundamental frequencies of test particle motion. The upper HFQPO corresponds to the azimuthal (Keplerian) orbital frequency, $\nu_{\rm U}=\nu_{\phi}$, while the lower HFQPO arises from the combination $\nu_{\rm L}=\nu_{\phi}-\nu_{r}$, where $\nu_{r}$ is the radial epicyclic frequency.  Notably $\nu_{\phi}$  is highly sensitive to the strong gravity region and to the detailed structure of the Kerr geometry, including frame dragging. In contrast, the radial epicyclic frequency $\nu_{r}$ is more susceptible to pressure forces, magnetic stresses, and other accretion disk effects. Consequently, within the RP model, the upper HFQPO is more directly influenced by the near horizon geometry in comparison to the lower HFQPO.  So it can be inferred that the corresponding upper QPO frequency vs temperature plot provides a more reliable geometric indicator. In the present case, this suggests that the source is more likely to lie on the LBH  branch.\\
We have repeated this analysis for all the other QPO models considered in this 
work, and in every case, we observe a similar qualitative behaviour.

\section{Summary and Concluding Remarks}\label{secVI}

In this work, we treated the QPO frequencies as a way to probe  the thermodynamic phase structure of the black hole.  Can QPOs be influenced by black hole phase transitions? Do they carry any signature of such transitions in their observational pattern? These were the central questions guiding our study.  We demonstrated that the behavior of both upper and lower QPO frequencies, when plotted against temperature reveals a clear imprint of the underlying phase structure.  Specifically, the frequencies trace out different black hole branches in a manner that reflects their respective thermodynamic stability. \\

We considered two different black hole system for our analysis,  the RN black hole and the Kerr black hole. Although the RN black hole is not astrophysically realistic,  it served as a useful toy model with a well-understood thermodynamic phase structure, providing a standard starting point for investigating black hole phase transitions. We aimed to isolate how geometric changes associated with thermodynamic transitions affect the QPO behavior. Our results showed consistent patterns across different thermodynamic phases, suggesting that QPOs may carry indirect signatures of black hole phase transitions. This appears to arise from the way geometric changes near the event horizon effects the particle motion.  For the Kerr black hole, we found that black hole spin significantly influenced the QPO–temperature relation, making the QPO frequencies more sensitive to thermodynamic changes in high-spin regimes. From the above study may be we can summarize as :  if the slope of the QPO–temperature curve is negative or zero, the thermodynamic phase is stable whereas if the slope is positive, the phase becomes unstable. Mathematically, this condition can be expressed as:
\begin{equation}
\begin{cases}
\dfrac{d\Omega_i}{dT} \leq 0, & \text{Stable phase}, \\[10pt]
\dfrac{d\Omega_i}{dT} > 0, & \text{Unstable phase}.
\end{cases}
\tag{1}
\end{equation}
Here $i$ refers to the upper or lower QPO frequencies.  To establish the above relation more rigorously, a broader range of black hole systems should be examined. With a refined mathematical framework, it may be possible to extract more information about the nature of this relation and the underlying physical mechanism. A detailed investigation in this direction is reserved for future work.\\

It should be noted that the present findings may be purely mathematical in nature and may not possess direct physical relevance, as the Hawking temperature is not yet an observable quantity. However, if a physical correspondence does exist, then QPOs could serve as an effective tool to probe the Hawking temperature and the associated black hole phase transitions. This may further suggest that QPOs are influenced not only by accretion-related processes but also by the underlying thermodynamic state of the black hole, which governs its spacetime geometry. \\

 We also compared the theoretical QPO frequencies obtained from Kerr black holes with the observed HFQPO data of several well known sources.  We analysed the behaviour of the upper and lower HFQPOs as functions of the Hawking temperature. For all sources considered, the upper HFQPOs intersect the LBH branch, whereas the lower HFQPOs fall exclusively on the SBH branch.  From our analysis we assume that the upper HFQPO points toward the LBH phase for the sources studied. Repeating the analysis across all QPO models considered in this work results the same qualitative behaviour.\\

 It is important to emphasise that the observed HFQPOs are originated in
the accretion disk, where additional physical processes can significantly
modify the frequency spectrum.  Hydrodynamic and magnetohydrodynamic effects such as pressure gradients, disk viscosity, magnetic stresses, turbulence, and disk corona interaction etc can
shift or modulate the QPOs.   Moreover, disk thickness, radiation pressure, resonant couplings, and
magnetically driven oscillations may also contribute to the measured QPO
peaks.   Therefore, while our analysis isolates the geometric contribution through
purely geodesic frequency calculations, the observed HFQPOs should be interpreted as
a combined outcome of both the spacetime geometry and the physics of
accretion flows.   QPOs observed in X-ray binaries have inspired a wide range of theoretical
interpretations.  One of the most widely used approaches is the relativistic precession (RP)
model, which associates the QPO frequencies with the precession of matter
orbiting in the strong gravitational field near the black hole.  
Epicyclic resonance (ER) models propose that HFQPOs arise from nonlinear
resonances between different oscillation mode typically the radial and
vertical epicyclic motionsin the accretion disk.  
Other frameworks, such as diskoseismology models, attribute QPOs to global
oscillation modes of the disk, while additional proposals involve warped disk
dynamics, magnetic field instabilities, or beat-frequency mechanisms.  
Although each model captures important aspects of disk dynamics, the true
physical origin of HFQPOs remains uncertain, and it is plausible that
different mechanisms may operate under different astrophysical conditions.
\\

In addition to these well known interpretations, our
work suggests that the spacetime geometry of the black
hole itself may also play a role in shaping QPO behavior.
Changes in the Hawking temperature are closely tied to
changes in the black hole’s geometry, particularly in the
near-horizon region. Since QPO frequencies are highly
sensitive to the structure of spacetime in the inner accretion zone, these geometric variations can significantly influence their behavior. In our interpretation, QPOs
are affected not only by accretion-related processes
but also by the underlying thermodynamic state of the
black hole, which governs its geometry. This is clearly
reflected in our results: as the black hole transitions
between different thermodynamic phases, the trend of
QPO frequencies with respect to temperature also shifts.
These findings suggest that QPOs may carry imprints
of the evolving geometry, and our work points to the
possibility that changes in black hole geometry could be
one of the contributing factors influencing QPO behavior.\\

These findings suggest that QPO frequencies may serve as powerful observational tools for probing the thermodynamic phase structure and stability of black holes. With the aid of available observational data, the imprints of phase transitions can be further substantiated, and the size of different black hole phases may be constrained. This presents a promising avenue for future research and warrants more detailed investigation.

\section*{Acknowledgements} 
\vspace{-0.2cm}
BH would like to thank DST-INSPIRE, Ministry of Science and Technology fellowship program, Govt. of India for awarding the DST/INSPIRE Fellowship[IF220255] for financial support. 	 We would also like to thank anonymous reviewer for useful suggestions and comments that improve our manuscript significantly.

\end{document}